\RequirePackage{tikz}
\usetikzlibrary{positioning,decorations.pathreplacing,calligraphy, calc, matrix}
\documentclass[11pt, a4]{article}% Default
\usepackage[top=1in, right=1.25in, left=1.25in, bottom=1.in, twoside]{geometry}
\usepackage{subcaption, amssymb}
\usepackage{fancyhdr, titling}
\usepackage[only,varcurlyvee]{stmaryrd}
\graphicspath{{./}{figures/}}
\newcommand{\pdiff}[2]{\frac{\partial#1}{\partial#2}}
\newcommand{\cgst}{C^T_{t_0}}
\newcommand{\nf}{\nabla \bm{F}^T_{t_0}}

\newcommand{\revadd}[1]{{#1}}
\newcommand{\fixme}[1]{{#1}}
\newcommand{\bmhead}[1]{\section*{\textbf{#1}}}

\usepackage{bm}
\usepackage[numbers]{natbib}
\usepackage{amsmath}
\usepackage{amssymb}
\usepackage{microtype}
\usepackage{graphics}
\usepackage{subcaption}
\usepackage{hyperref}
\usepackage{url}
\usepackage{booktabs}
\usepackage{rotating}

\title{Three-dimensional Lagrangian Coherent Structures in the Elliptic-Restricted Three-body Problem}
\author{Tyler, Jack\thanks{Corresponding author, PhD Candidate, Astronautics Research Group, University of Southampton, University Road, Southampton, SO17 1BJ. \url{jack.tyler@soton.ac.uk}}\, and Wittig, Alexander\thanks{Associate Professor, Astronautics Research Group, University of Southampton, University Road, Southampton, SO17 1BJ}}

\begin{document}
\date{}
\pagestyle{fancy}

% \author*[1]{\fnm{Jack} \sur{Tyler}}\email{jack.tyler@soton.ac.uk}
% \author[2]{\fnm{Alexander} \sur{Wittig}}\email{a.wittig@soton.ac.uk}

% \affil[1]{PhD Candidate, \orgdiv{Astronautics Research Group}, \orgname{University of Southampton}, \orgaddress{\street{University Road}, \city{Southampton}, \postcode{SO17 1BJ}, \country{United Kingdom}}}
% \affil[2]{Associate Professor, \orgdiv{Astronautics Research Group}, \orgname{University of Southampton}, \orgaddress{\street{University Road}, \city{Southampton}, \postcode{SO17 1BJ}, \country{United Kingdom}}}

% %%==================================%%
% %% sample for unstructured abstract %%
% %%==================================%%

% \keywords{Dynamical Systems, Lagrangian Coherent Structures, Differential Algebra, Three-body Problem}



\maketitle

\begin{abstract}
    In the preliminary design of space missions it can be useful to identify regions of dynamics that drive the system's behaviour or separate qualitatively different dynamics. The Lagrangian Coherent Structure (LCS) has been widely used in the analysis of dynamical systems, and generalises the concept of the stable and unstable manifolds to systems with arbitrary time-dependence. However, the use of three-dimensional LCS in astrodynamics has thus far been limited. This paper presents the application of a new numerical method introduced by the authors, DA-LCS, to astrodynamics systems using the Elliptic-Restricted Three-body Problem (ER3BP) as a test case. \revadd{We are able to construct the full, three-dimensional LCS associated with the Sun-Mars ER3BP directly from the variational theory of LCS even for  numerically challenging initial conditions.} The LCS is analysed in detail, showing how it in this case separates regions of qualitatively different behaviour without any \textit{a priori} knowledge. The paper then studies the effect of integration time and the parameterisation of the initial condition on the LCS found. \revadd{We highlight how round-off errors arise from limits of floating-point arithmetic in the most challenging test cases and provide mitigating strategies for avoiding these errors practically.}
\end{abstract}

\section{Introduction}\label{sec:introduction}

In designing space missions, it can be useful to identify regions of dynamics which drive system behaviour or separate regions of qualitatively different dynamics. In time-independent systems, classical dynamical systems theory identifies the invariant manifolds as separators of qualitatively different flow in phase space, which are found by studying the system's behaviour over long time scales \citep{Meiss1992SymplecticTransport}. 
However, in systems with arbitrary time-dependence, such structures do not always exist. Instead, one often studies the behaviour of the system over some practical time period of interest.

Several different methods have been used to attempt to identify analogous structures to the invariant manifolds in systems with arbitrary time-dependence. The Finite-time Lyapunov Exponent (FTLE) is a metric that quantifies the separation of infinitesimally-close trajectories over time \citep{Shadden2005DefinitionFlows}. One might expect that high FTLE values signal the presence of separatrices in the flow, but studies have shown that the FTLE does not necessarily indicate the presence of separatrices \citep{Haller2011LagrangianExponent} and in some cases the hyperbolicity of local dynamics can affect the FTLE value \citep{Gondelach2019OnOrbits}. In astrodynamics research, the FTLE (and the closely-related Fast Lyapunov Indicator) has been used to study chaotic astrodynamics systems and profile behaviour since 1997 \citep{Froeschle1997TheBelt, Villac2008UsingEnvironments, Lega2016TheoryMethod}.

Since the FTLE does not necessarily highlight separatrices, and because it requires the computationally-expensive evaluation of derivatives, alternative heuristic measures have been investigated. The Lagrangian Descriptor (LD) has seen use in astrodynamics to attempt to identify regions of qualitatively different behaviour \citep{Quinci2022, Raffa2022}. The LD involves the integration of a positive, bounded quantity along a given trajectory. While for some choices of integrand, discontinuities in the LD scalar field can identify separatrices between regions of qualitatively different behaviour, the selection of the integrand can greatly change the result and must be selected for each particular flow \citep{Mancho2013}. Moreover, being a heuristic indicator there is no rigorous mathematical relation between the LD and any separatrices.

In 2000, \citet{Haller2000LagrangianTurbulence} defined the Lagrangian Coherent Structure (LCS), surfaces that exert significant influence on nearby trajectories.
A particular type of LCS, the hyperbolic LCS, is a surface that acts as the locally most repelling or attracting region of flow, playing an analogous role to the unstable and stable invariant manifolds respectively. Several equivalent methods for determining LCS exist \citep{Hadjighasem2017ADetection}; original definitions of the hyperbolic LCS were made with respect to ridges of the FTLE \citep{Shadden2005DefinitionFlows}, but it was later discovered that these can only be related rigorously to LCS under certain additional variational conditions \citep{Haller2011AStructures}. 

A more recent formulation in \citet{Blazevski2014HyperbolicFlows} presented a method to determine LCS in three-dimensions from their variational theory. This method studies the eigenvectors of the right Cauchy-Green Strain Tensor (CGST), which quantifies the deformation of a flow over a given trajectory, and seeks intersections of the repelling and attracting LCS with a set of hyperplanes, deemed `strainlines' and `stretchlines'. This approach objectively and without any \textit{a priori} knowledge determines LCS in systems with arbitrary time-dependence, while respecting their original mathematical definition as the locally most repelling or attracting surfaces.

The LCS has been used in a wide array of fields for profiling the behaviour of dynamical systems \citep{Nolan2020PollutionStructures, Rutherford2012LagrangianIntensification, Tu2019InvestigationStructures, Fiorentino2012UsingQuality}. LCS derived from the FTLE field has also been well-studied in astrodynamics. \citet{Gawlik2009LagrangianProblem} investigated LCS and their representations in the planar Elliptic-Restricted Three-body Problem (ER3BP), and \citet{Short2011LagrangianProblem} studied the use of LCS in various map representations in the planar ER3BP. \citet{Qingyu2020LagrangianProblem} extracted the ridges of the FTLE field using a Particle Swarm optimisation algorithm in the planar and hyperbolic restricted three-body problem. Both \citet{Short2015StretchingProblems} and \citet{inproceedings} went on to analyse the behaviour of particles in the Sun-Earth-Moon bicircular problem using regions of high FTLE as initial conditions for strainlines and stretchlines.

Two MSc Theses, \citet{RosRoca2015ComputationBoundaries} and \citet{Parkash2019ApplicationTrajectories}, applied the full variational theory of LCS to inform the identification of ballistic capture orbits in the planar ER3BP \citep{Luo2015}. However, significant numerical difficulties and computational complexities in the numerical method were encountered which prevented the determination of the full LCS.
This numerical difficulty comes from the correct approximation of $\cgst$ and the derivatives of its eigenvalues and eigenvectors. Near the repelling LCS, maximal repulsion is necessarily expected, and thus the CGST becomes large and potentially ill-conditioned. Moreover, the eigenvectors of the CGST must be determined accurately to correctly track along the structure of the LCS. The approximation of divided differences is often used to determine the derivatives of the flow and the derivatives of the eigenvectors and eigenvalues of $\cgst$, but the grid over which to approximate the derivative must be chosen carefully, often without any \textit{a priori} insight \citep{Farazmand2012ComputingTheory}. \citet{Sanchez-Martin2018FromModels}, who were able to compute a two-dimensional LCS in a galactic bar model, remarked that it was `crucial' to correctly choose the grid-sizes used for approximating derivatives to obtain a valid result.

A LCS is also computationally expensive to compute, especially in three-dimensions \citep{Farazmand2012ComputingTheory}. To reduce the computational expense, several different approaches have been considered. \citet{Lipinski2010AStructures} used a ridge-tracking algorithm that prevented the computation of trajectories away from the LCS, instead constructing each portion of the structure iteratively.
Many different approaches to adaptive mesh refinement have also been used to decrease run-times and improve accuracy near complex features in the flow \citep{Miron2012AnisotropicStructures}.
These methods work in tandem with standard approaches to program parallelism and high-performance computing techniques. \citet{Lin2017GPU-acceleratedRegimes} published work that details the parallelism of LCS computation using the FTLE field on Graphics Processing Units (GPUs) using NVIDIA's CUDA programming paradigm.

To address both the numerical difficulties and computational expense, in \citet{Tyler2022} the authors recently outlined an improved numerical method, DA-LCS, for computing LCS efficiently and automatically in three-dimensions, without the need to manually determine or approximate derivatives. DA-LCS can produce insight where traditional methods of approximating derivatives fails to produce any. Differential Algebra (DA), which provides the full set of operations to store and manipulate Taylor polynomials in a computer environment \citep{Berz1999, Rasotto2016, Massari2018DifferentialApplications}, is used to compute derivatives of flows of the dynamical system and the eigenvectors of $\cgst$ accurate to machine precision. 

This paper will advance the use of LCS in astrodynamics by using this improved numerical method to compute LCS in the Elliptic-Restricted Three-body Problem (ER3BP), with applications to arbitrary astrodynamics systems. We will first introduce the numerical method, and then \revadd{apply it to a numerically challenging test case to show how it can be used to accurately and automatically determine LCS in astrodynamics systems}. We analyse the LCS in detail, including highlighting how the LCS acts in this case as a separatrix between qualitatively different behaviours. The paper then examines how the parameterisation of the initial condition affects the LCS found, exploiting the fact that DA allows us to automatically compute derivatives of parameterisations without further effort. From this investigation, we highlight a numerical pitfall that must be considered when computing LCS practically and provide guiding considerations on how to best choose how to define the initial conditions in practice.

\section{Background}

\subsection{Differential Algebra}\label{sec:differentialalgebra}

In the following, we give a very brief introduction to Differential Algebra. For a comprehensive treatment, the reader is referred to the literature \citep{Berz1999}.

Differential Algebra (DA) was originally introduced to determine high-order transfer maps for particle accelerator systems \cite{Berz1987}, and provides the tools necessary to compute the derivatives of functions within a computer environment \cite{Cavenago2017On-boardNavigation, Berz1999}.
DA constructs a Taylor series representation of an arbitrary map, and has seen widespread use in the study of non-linearities \cite{Makino1996RemainderApplications, Makino1998RigorousAccelerators, DiMauro2015NonlinearAlgebra}, the management of uncertainties \cite{Wittig2014, Wittig2015PropagationSplitting, Massari2017, Armellin2010AsteroidApophis}, and as a tool for automatic differentiation \cite{automaticdifferentiation, Massari2018DifferentialApplications}. More generally, DA has also been applied to numerous other astrodynamics problems, such as the Two-Point Boundary Value Problem; spacecraft guidance and state estimation; trajectory optimisation; and orbital conjunctions \cite{Lizia2008, DiLizia2012, Zazzera2012, Armellin2012AnModels, Cavenago2018}.

A full introduction to the inner workings of DA is beyond the scope of this paper (see \citet{Berz1987}), but as an example to introduce DA, consider two real numbers $a$ and $b \in \mathbb{R}$. The approximation to $a$ and $b$ in a computational environment is their floating-point representation $\bar{a},\,\bar{b} \in \mathbb{F}$, which essentially stores a set number of digits of its binary expansion. Any operation defined in $\mathbb{R}$, $\Box$, has a corresponding operation in $\mathbb{F},\,\boxtimes$, defined such that the result is another floating-point approximation of the operation on the real numbers $a$ and $b$, i.e. $\bar{a}\times\bar{b}$ commutes with the floating-point representation of $a\times b$, $\overline{a\times b}$.

Similarly, now consider two functions, $c$ and $d$, which are sufficiently smooth, $k-$differentiable functions of $n$ variables: $c,\,d : \mathbb{R}^n \rightarrow \mathbb{R}$. In the DA framework, a computer operates on the multivariate Taylor expansion of $c$ and $d$, $[c]$ and $[d]$, with corresponding operations to those defined in the real function space, such that the operation of $[c]\cdot[d]$ commutes with the DA representation of the product $[c\cdot d]$.

Differential Algebra provides a full set of elementary operations to efficiently operate on these multivariate expansions \citep{Wittig2010RigorousSurfaces}. This includes operations for common intrinsic functions such as division, square roots, trigonometric functions, and exponentials, as well as operations for differentiation and integration \cite{Wittig2015PropagationSplitting}. With these, any arbitrarily complicated function containing these operations can be coded and evaluated in DA.

Because we are constructing a Taylor expansion to arbitrary order, and we have access to operators for differentiation and integration, DA can be viewed as a form of automatic differentiation. Conceptually, this provides similar functionality as, for example, symbolic math libraries in Matlab or Python which compute algebraic expressions that can be differentiated analytically. But instead of representing functions as algebraic expressions, DA represents them as the coefficients of the Taylor expansion. This makes DA computationally more efficient, especially for complicated functions.

A key advantage of this automatic differentiation is that the derivatives of solutions of ODEs with respect to their initial conditions can be taken automatically, a concept known as flow expansion \citep{Cavenago2018, Zazzera2012, Perez-Palau2015ToolsTransport}. Practically, this is achieved by replacing the floating-point arithmetic in standard numerical ODE integration schemes by the DA arithmetic. In modern programming languages with operator overloading such as C++, this process is straightforward.

The Differential Algebra Computational Engine (DACE) is used to implement DA in the program \citep{Massari2018DifferentialApplications}. The numerical integration is performed using the 7/8\textsuperscript{th} Dormand-Prince Runge-Kutta integration as part of the Boost C++ library \citep{ahnertOdeintSolvingOrdinary2011}. Custom norms are defined to connect the step-size control algorithms in the numerical integrators in Boost and DA objects: since evaluating a vector norm in DA yields another DA expression, for the purposes of Boost we define the norm of a DA object to map it into the non-negative real numbers. The norm of a DA object in this paper is the largest absolute value of any coefficient of the expansion in any order.

\subsection{The Cauchy-Green Strain Tensor and Lagrangian Coherent Structures}\label{sec:cgstlcs}

The numerical method for computing LCS using DA, summarising the main results of \citet{Tyler2022}, is now introduced. For a more comprehensive introduction to the mathematical formulation, the reader is referred to \citet{Blazevski2014HyperbolicFlows}.

We study the behaviour of a dynamical system
\begin{equation}\label{eq:dynamicalsystemdefinition}
    \dot{\bm{x}} = f\left(\bm{x},~t\right), \bm{x} \in D \subset \mathbb{R}^n,\,t \in \left[t_0,\,T\right]
\end{equation}
where $f$ is a smooth vector field considered from time $t_0$ to time $T$. Denoting a trajectory of the system starting at position $\bm{x}_0$ at time $t_0$ up to a time $T$ as $\bm{x}\left(t_0,\,\bm{x}_0;\,T\right)$, the flow map of Equation \ref{eq:dynamicalsystemdefinition} is given by 
\begin{equation}
    \bm{F}^T_{t_0} : \begin{cases}
        D \rightarrow D \\
        \bm{x}_0 \mapsto \bm{x}\left(t_0,\,\bm{x}_0;\,T\right)
    \end{cases}
\end{equation}
which is assumed to be at least twice continuously differentiable. 
The right Cauchy-Green Strain Tensor (CGST) $\cgst$ is defined by the Jacobian of this flow map, $\nf$, and describes the deformation of the flow at the end of the given trajectory
\begin{equation}\label{eq:cgstflowmap}
    \cgst = \left(\nabla \bm{F}^T_{t_0}\right)^\top \left(\nabla \bm{F}^T_{t_0}\right)
\end{equation}
with $^\top$ the matrix transpose. $\cgst$ is positive-definite and symmetric, with real eigenvalues $0 < \lambda_1 \leq \lambda_2 \leq \dots \leq \lambda_n$ and associated real eigenvectors $\bm{\zeta}_1, \bm{\zeta}_2, \dots, \bm{\zeta}_n$.
Recalling that the full, three-dimensional LCS is the surface that is locally maximally repelling or attractive over a time interval $\left[t_0,\,T\right]$, this surface is necessarily everywhere orthogonal to either $\bm{\zeta}_n$ or $\bm{\zeta}_1$, respectively.

The LCS is constructed from its intersection with  series of reference hyperplanes $\mathcal{S}$. For repelling LCS, the intersections are termed \textit{reduced strainlines}; for attracting LCS, the intersections are \textit{reduced stretchlines}. In the following, we show the mathematical formulation for repulsive LCS, whose structure is derived from the dominant eigenvector $\bm{\zeta}_n$. A similar procedure applies to $\bm{\zeta}_1$ to obtain attracting LCS.

We first sample points on each hyperplane in $\mathcal{S}$ on a uniformly-spaced grid and compute the helicity $H_{\bm{\zeta}_n}$ at each point
\begin{equation}\label{eq:helicity}
    H_{\bm{\zeta}_n} = \langle\nabla\times\bm{\zeta}_n,\,\bm{\zeta}_n\rangle.
\end{equation}
Mathematically, points that satisfy $H_{\bm{\zeta}_n} = 0$ are maximisers of repulsion. Numerically, this condition is relaxed such that points with helicity below some threshold $\alpha > 0$ are seed points for an ODE to propagate the strainline forward. The ODE is tangent to the reduced strainline and tracks within the hyperplane. In discretised form, the ODE is
\begin{equation}\label{eq:strainlineode}
    \bm{s}_i^\prime = \text{sign}\left(\bm{\zeta}_{i,\,n}\cdot\bm{\zeta}_{i-1,\,n}\right)\hat{\bm{n}}_{\mathcal{S}_i}\times\bm{\zeta}_{i,\,n}
\end{equation}
where $\bm{s}_i$ is the $i-$th point on the strainline and the term $\bm{\zeta}_{i,\,n}\cdot\bm{\zeta}_{i-1,\,n}$ is introduced to enforce continuity in the vector field by selecting the direction most closely aligned with the previous tangent vector. The unit vector $\hat{\bm{n}}_\mathcal{S}$ is the normal to the hyperplane at that point. The numerical integration of the ODE along the strainline continues until the sum of the helicity at each $\bm{s}_i$ divided by the number of steps performed ($i$) rises above $\alpha$.

The trajectories of Equation \ref{eq:strainlineode} are segments of strainlines forming the LCS. However, since different initial points can belong to the same strainline, the trajectories often overlap. They must, therefore, be filtered to provide a single, continuous curve. If this occurs, the Fr\'echet distance defined in \citet{Driemel2016} is used to filter strainline segments.

This analysis is repeated for each of the hyperplanes in $\mathcal{S}$. The strainlines forming part of the LCS on each hyperplane are then interpolated to produce the full 3D structure of the LCS.

In \citet{Tyler2022}, we show how we use DA to compute Equations \ref{eq:helicity} and \ref{eq:strainlineode} to high accuracy. We first use DA to construct an expansion of $\cgst$ with respect to the initial parameterisation using standard flow expansion techniques \citep{Wittig2012RigorousManifolds, Armellin2010AsteroidApophis, Lunghi2018AtmosphericLanding, Zazzera2012}, accurate to machine precision. This provides accurate determination of $\bm{\zeta}_n$ for use in Equation \ref{eq:strainlineode}. Using a novel modified power law method, another expansion of $\bm{\zeta}_n$ with respect to the initial parameterisation is then constructed to obtain the seed points in Equation \ref{eq:helicity} to high accuracy. A key advantage of using DA is that the relevant quantities for the LCS are computed completely automatically, without the need to adjust grid sizes, or manually derive and implement explicit derivatives or variational equations.

\section{Dynamical Models}

\subsection{Equations of Motion}

This paper studies motion in the Sun-Mars Elliptic-Restricted Three-body Problem (ER3BP). A generalisation of the $n-$body problem for $n=3$, the ER3BP tracks the motion of an infinitesimally small mass $m_3$ under the gravitational influence of two far larger masses $m_1$ (Sun) and $m_2$ (Mars), $m_1 > m_2 >> m_3$, which orbit each other in an ellipse. The system is parameterised by the mass parameter $\mu = m_2 / \left(m_1 + m_2\right)$ and the eccentricity of the ellipse of the orbit of $m_2$ about $m_1$, $e_p$. The special case of $e_p = 0$ yields an autonomous system where the classical invariant manifolds partition phase space, but for $e_p\neq 0$ these are not as easily defined. For the Sun-Mars case, $\mu = 3.227\times 10^{-7}$ and $e_p = 0.0935$.

The equations of motion for this system for the motion of $m_3$ are defined in a rotating-pulsating reference frame centred on the barycentre of $m_1$ and $m_2$ where the $m_1$-$m_2$ distance is normalised to unity. For a position vector $\bm{x} = \left(x, y, z\right)$ with velocity $\bm{x}^\prime = \left(x^\prime,\,y^\prime,\,z^\prime\right)$, the equations of motion are
\begin{eqnarray}\label{eq:er3bpeomstart}
x^{\prime\prime} &=& 2y^\prime + \pdiff{\Omega}{x}\\
y^{\prime\prime} &=& -2x^\prime + \pdiff{\Omega}{y}\\
z^{\prime\prime} &=& \pdiff{\Omega}{z}\label{eq:er3bpeomend}
\end{eqnarray}
where
\begin{equation}
    \Omega = \frac{1}{1 + e_p\cos\nu}\left[ \frac{1}{2}\left(x^2 + y^2 - z^2 e\cos\nu\right) + \frac{\mu}{r_1} + \frac{1-\mu}{r_2}\right]
\end{equation}
and
\begin{eqnarray}
    r_1 &=& \sqrt{\left(x + \mu\right)^2 + y^2 + z^2}\\
    r_2 &=& \sqrt{\left(x - 1 + \mu\right)^2 + y^2 + z^2}.
\end{eqnarray}
The variable $\nu$ is the true anomaly, the angle $m_2$ makes with respect to $m_1$ in an inertial Cartesian coordinate system, and is used as the independent variable. We use the notation $\dot{\Box}$ to represent derivatives with respect to time, and $\Box^\prime$ to represent derivatives with respect to the true anomaly $\nu$.

\subsection{Definition of initial conditions}\label{sec:initialconditionstransfermap}

The algorithm for computing LCS introduced in Section \ref{sec:cgstlcs} is explicitly designed for a CGST that is $3\times 3$ in dimension and hence represents the behaviour of a three-dimensional system. However, the ER3BP lives in a phase space in $\mathbb{R}^6$. To simplify visualisation and enable analysis of the ER3BP, we embed a three-dimensional sub-manifold in the six-dimensional phase space on which we then compute the LCS. We arrive at this embedding by first parameterising a region of interest in three dimensions in a non-rotating frame about Mars, and then defining a transformation $\Psi: \mathbb{R}^3\mapsto\mathbb{R}^6$ to complete the full phase space and transform the initial condition into the ER3BP rotating-pulsating frame. After propagation under the ER3BP equations of motion, the inverse transformation $\Pi: \mathbb{R}^6\mapsto\mathbb{R}^3$ takes the final condition back into the original parameterisation. This process is shown below for an arbitrary starting parameterisation $\left(\kappa,\beta,\,\varsigma\right) \in \mathbb{R}^3$:

\begin{center}
{\centering
\begin{tikzpicture}
\centering
\matrix (m) [matrix of math nodes, row sep=3.5em, column sep=3.5em, nodes={anchor=center}]
 { \left(x,\,y,\,z,\,x^\prime,\,y^\prime,\,z^\prime\right)& \left(x,\,y,\,z,\,x^\prime,\,y^\prime,\,z^\prime\right)_F \\
  \left(\kappa,\,\beta,\,\varsigma\right) & \left(\kappa,\,\beta,\,\varsigma\right)_F \\};
\path[-stealth]
 (m-1-1) edge node [above] {$f$} (m-1-2)
 (m-2-1) edge node [above] {$\Psi\circ f\circ \Pi$} (m-2-2)
 (m-2-1) edge node [right] {$\Psi$} (m-1-1)
 (m-1-2) edge node [right] {$\Pi$} (m-2-2);
\end{tikzpicture}
}
\end{center}

\noindent where the subscript $_F$ implies final states. In the non-rotating frame, the distance unit is chosen such that the Sun-Mars distance is unity; the rationale behind this is given in Section \ref{sec:inversion}. For more information on the conversion between the non-rotating and rotating reference frames, the reader is referred to \citet{Szebehely1968}. \revadd{We refer to $\left(\kappa,\,\beta,\,\varsigma\right)$ as the parameterisation of the manifold and the map $\Psi\cdot f\cdot \Pi$ as the transformation.}

Three pairings of $\Psi$ and $\Pi$ are explored in this paper and are elaborated in the following. A key advantage of DA-LCS is that for all $\Psi$ and $\Pi$ given below, provided the operations are coded using DA arithmetic the derivatives are computed fully automatically and so there is no need to manually derive or implement derivatives. This is particularly useful when considering that these transformations are in some cases non-trivial and otherwise very difficult to differentiate manually.

\subsubsection{Orbital elements}

In this transformation, the initial position and velocity are given by standard Keplerian orbital elements in the non-rotating frame, following their use in the literature to define initial conditions for ballistic capture in three dimensions \citep{Luo2014ConstructingModel}. 

The sub-manifold is parameterised by the radius of periapsis $r_p$, inclination $i$ and argument of periapsis $\omega$ of a purely Keplerian orbit around Mars. The hyperplanes $\mathcal{S}$ are given by fixed values of $i$ (Figure \ref{f:mappings:orbital_elements}). The full element set is obtained by fixing the remaining orbital elements $e = 0.95$, $\Omega = 118^\circ$, $M = 0$ based on values suggested in the literature to facilitate ballistic capture \citep{Luo2015}. \revadd{This choice 
of $\Omega$ maximises `interesting' dynamical behaviour across all inclinations, as the high eccentricity leads to orbits which escape into heliocentric orbits rapidly. This presents a numerically very challenging test case to compute the LCS for.}

The full orbital element set at each point is converted to equivalent Cartesian position and velocity in the non-rotating frame (subscript $_I$), which is then transformed into the ER3BP rotating-pulsating frame (subscript $_R$) and propagated under the full equations of motion (Equation \ref{eq:psioes}).

\begin{gather}
  \Psi_{OE}: \begin{Bmatrix}\label{eq:psioes}
    r_p\\i\\\omega
    \end{Bmatrix}_I \mapsto
    \begin{Bmatrix}
        r_p\\e\\i\\\Omega\\\omega\\M
    \end{Bmatrix}_I \mapsto
    \begin{Bmatrix}
    x\\y\\z\\\dot{x}\\\dot{y}\\\dot{z}
    \end{Bmatrix}_I \mapsto
    \begin{Bmatrix}
    x\\y\\z\\x^\prime\\y^\prime\\z^\prime
    \end{Bmatrix}_R\\
    \Pi_{OE}: \begin{Bmatrix}\label{eq:pioes}
    x\\y\\z\\x^\prime\\y^\prime\\z^\prime
    \end{Bmatrix}_R \mapsto
    \begin{Bmatrix}
    x\\y\\z\\\dot{x}\\\dot{y}\\\dot{z}
    \end{Bmatrix}_I \mapsto
    \begin{Bmatrix}
        r_p\\e\\i\\\Omega\\\omega\\M
    \end{Bmatrix}_I \mapsto
    \begin{Bmatrix}
        r_p\\i\\\omega
    \end{Bmatrix}_I.
\end{gather}

The inverse transformation $\Pi$ is as in Equation \ref{eq:pioes}, and takes the final condition in the rotating-pulsating frame of the ER3BP, transforms it back into the non-rotating frame and converts this position and velocity into its equivalent instantaneous orbital elements. The radius of periapsis, inclination and argument of periapsis of the final condition are then isolated from the full set for computing $\cgst$.

The conversion to-and-from orbital elements and Cartesian position is performed using standard conversion algorithms presented in \citet{Curtis2009} modified to support DA arithmetic.

\subsubsection{Spherical coordinates}\label{sec:sphparam}

In this transformation, the sub-manifold is parameterised using spherical coordinates $\bm{\psi} = \left(\rho,\,\theta,\,\phi\right)$ where hyperplanes are given by fixed $\phi$ (Figure \ref{f:mappings:spherical_coordinates}). This is a natural extension of the use of polar coordinates to compute ballistic capture sets in two dimensions in the literature \citep{Hyeraci2010MethodProblem}. The phase space is completed by uniquely associating a velocity with each point in physical space. 

\revadd{This velocity is defined such that it is the velocity a point in the orbital element transformation would have at the same Cartesian position. To ensure a valid comparison and avoid leaking information from one set of coordinates to the other, we need to define this velocity independently of any of the original orbital elements. For the initial Cartesian position $\bm{x} = \left(x,\,y,\,z\right)$ corresponding to $\bm{\psi}$} we have
\begin{align}
    x =& \rho\cos\theta\sin\phi \\
    y =& \rho\sin\theta\sin\phi \\
    z =& \rho\cos\phi
\end{align}
and can define the velocity in the non-rotating frame at this point $\bm{v}\left(\bm{x}\right)$ as
\begin{equation}\label{eq:velocitydefinition}
    \bm{v}\left(\bm{x}\right) = \sqrt{\text{GM}\frac{1+e}{\rho^3}}\left[\begin{pmatrix}
        x\\y\\z
    \end{pmatrix}\times \hat{\bm{n}}\left(\bm{x}\right)\right]
\end{equation}
where $\text{GM}$ is the standard gravitational parameter of Mars in the non-rotating frame, and $e$ has the same meaning and values as before. The quantity $\hat{\bm{n}}$ is the normal to the orbital plane at that point and can be computed by considering the vector that points along the line of nodes $\hat{\bm{\Omega}} = \left(\cos\Omega,\,\sin\Omega,\,0\right)^\top$
\begin{equation}
    \hat{\bm{n}}\left(\bm{x}\right) = \text{sign}\left(z\right)\frac{\hat{\bm{\Omega}}\times \bm{x}}{\lvert\lvert \hat{\bm{\Omega}}\times \bm{x} \rvert\rvert}.
\end{equation}
The value of $\Omega$ used is the same as in the orbital elements transformation, and the $\text{sign}\left(z\right)$ term is required to ensure the correct orientation of the unit normal. 
\revadd{Equation \ref{eq:velocitydefinition} is well defined everywhere except on the line of nodes, where $\hat{\bm{\Omega}}\times\bm{x} = \bm{0}$. Since there is no one unique inclination passing through these points, additional information from the orbital elements transformation would be required to attach an equivalent velocity at these points. The reference hyperplanes should thus be selected or sampled in such a way to avoid these ill-defined points.}

Once the full phase space is complete, it is converted into the ER3BP rotating-pulsating frame and propagated under the equations of motion
\begin{equation}
     \Psi_S: \begin{Bmatrix}\label{eq:psisph}
    \rho\\\theta\\\phi
    \end{Bmatrix}_I\mapsto
    \begin{Bmatrix}
        x\\y\\z
    \end{Bmatrix}_I \mapsto
    \begin{Bmatrix}
    x\\y\\z\\\dot{x}\\\dot{y}\\\dot{z}
    \end{Bmatrix}_I \mapsto
    \begin{Bmatrix}
    x\\y\\z\\x^\prime\\y^\prime\\z^\prime
    \end{Bmatrix}_R.\\
\end{equation}
The spatial dimensions are then converted back into spherical coordinates in the non-rotating frame to compute $\cgst$
\begin{equation}
\Pi_S: \begin{Bmatrix}\label{eq:pisph}
    x\\y\\z\\x^\prime\\y^\prime\\z^\prime
    \end{Bmatrix}_R \mapsto
    \begin{Bmatrix}
    x\\y\\z\\\dot{x}\\\dot{y}\\\dot{z}
    \end{Bmatrix}_I \mapsto
    \begin{Bmatrix}
        x\\y\\z
    \end{Bmatrix}_I \mapsto
    \begin{Bmatrix}
        \rho\\\theta\\\phi
    \end{Bmatrix}_I.
\end{equation}

\subsubsection{Cartesian coordinates}\label{sec:cartesianparam}

\begin{figure}
\centering
\begin{subfigure}{0.8\textwidth}
    \centering
    \def\svgwidth{\linewidth}
    \renewcommand{\theta}{M}
    \includegraphics[width=\linewidth]{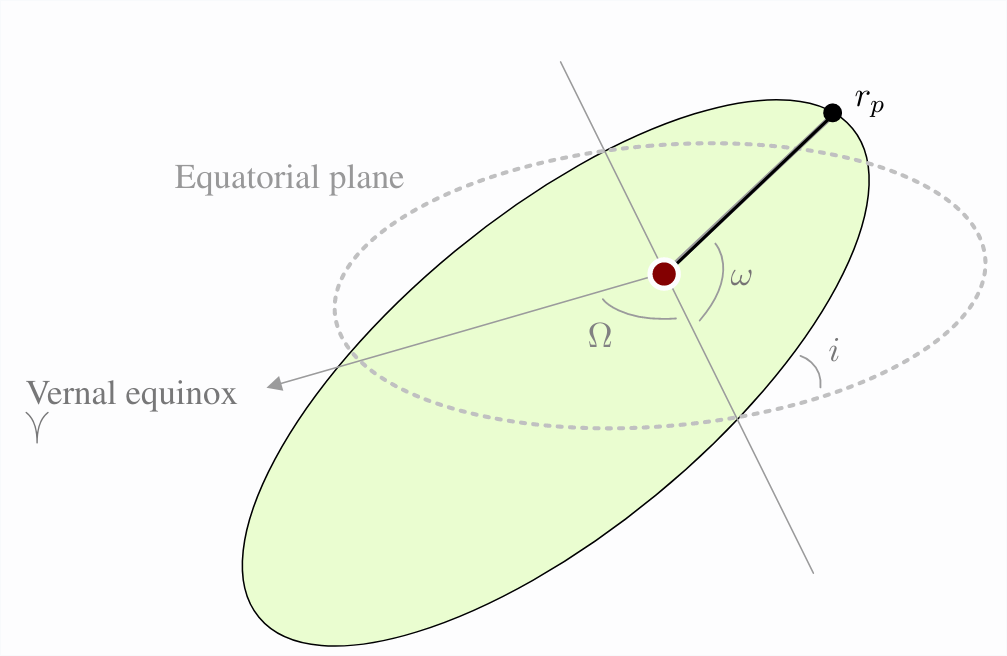}
    \caption{Graphical depiction of the parameterisation of an initial condition using standard Keplerian elements. The light green plane represents an example hyperplane of fixed $i$.}
    \label{f:mappings:orbital_elements}
\end{subfigure}
\begin{subfigure}{0.45\textwidth}
    \centering
    \def\svgwidth{\linewidth}
    %% Creator: Inkscape 1.2.1 (9c6d41e4, 2022-07-14), www.inkscape.org
%% Accompanies image file 'spherical_coords.pdf' (pdf, eps, ps)
\begingroup%
  \makeatletter%
  \providecommand\color[2][]{%
    \errmessage{(Inkscape) Color is used for the text in Inkscape, but the package 'color.sty' is not loaded}%
    \renewcommand\color[2][]{}%
  }%
  \providecommand\transparent[1]{%
    \errmessage{(Inkscape) Transparency is used (non-zero) for the text in Inkscape, but the package 'transparent.sty' is not loaded}%
    \renewcommand\transparent[1]{}%
  }%
  \providecommand\rotatebox[2]{#2}%
  \newcommand*\fsize{\dimexpr\f@size pt\relax}%
  \newcommand*\lineheight[1]{\fontsize{\fsize}{#1\fsize}\selectfont}%
  \ifx\svgwidth\undefined%
    \setlength{\unitlength}{595.27559055bp}%
    \ifx\svgscale\undefined%
      \relax%
    \else%
      \setlength{\unitlength}{\unitlength * \real{\svgscale}}%
    \fi%
  \else%
    \setlength{\unitlength}{\svgwidth}%
  \fi%
  \global\let\svgwidth\undefined%
  \global\let\svgscale\undefined%
  \makeatother%
  \begin{picture}(1,0.7047619)%
    \lineheight{1}%
    \setlength\tabcolsep{0pt}%
    \put(0,0){\includegraphics[width=\unitlength,page=1]{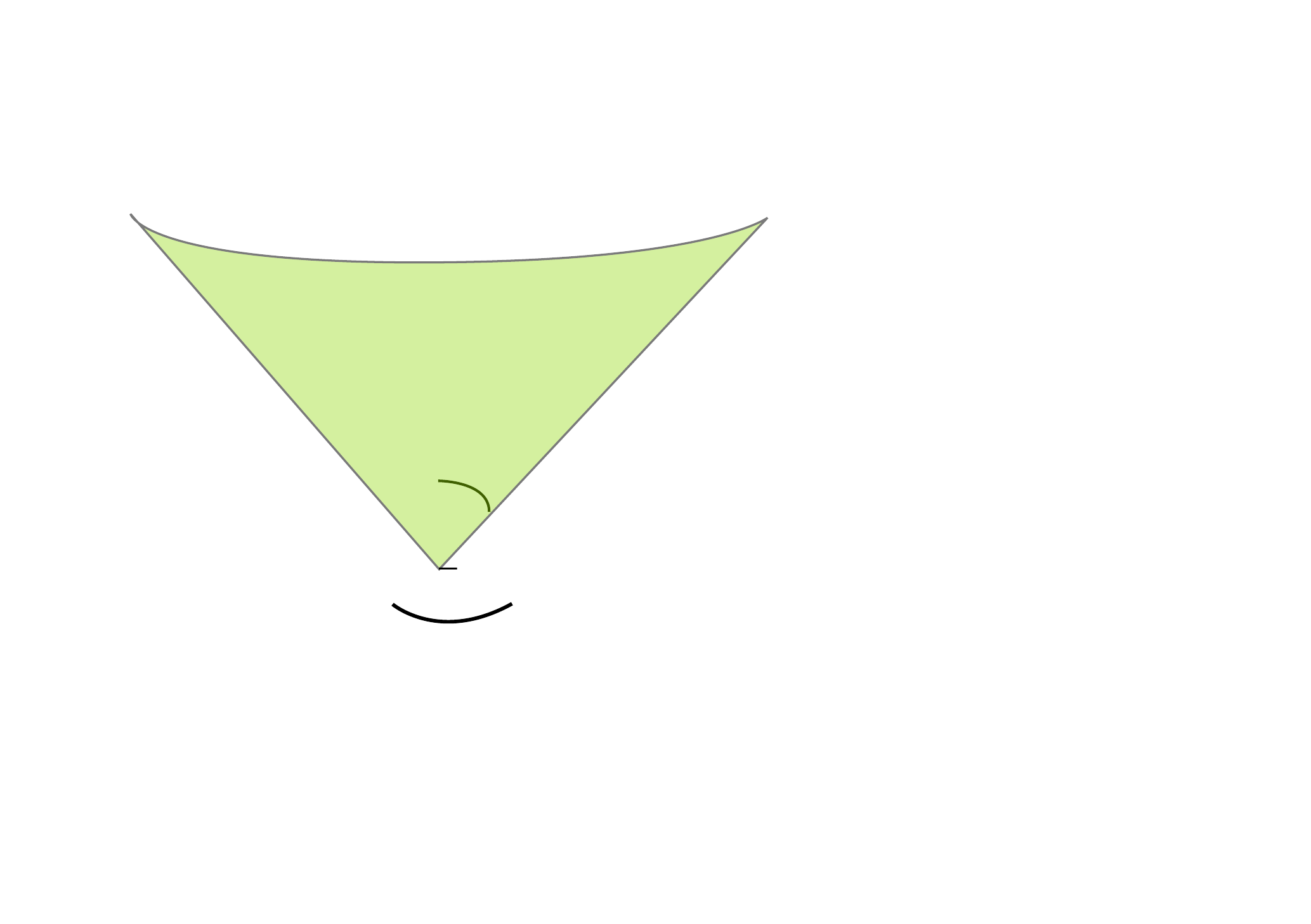}}%
    \put(0.36269815,0.35455241){\color[rgb]{0,0,0}\makebox(0,0)[lt]{\lineheight{1.25}\smash{\begin{tabular}[t]{l}$\phi$\end{tabular}}}}%
    \put(0.31699153,0.17636915){\color[rgb]{0,0,0}\makebox(0,0)[lt]{\lineheight{1.25}\smash{\begin{tabular}[t]{l}$\theta$\end{tabular}}}}%
    \put(0.53234081,0.4330974){\color[rgb]{0,0,0}\makebox(0,0)[lt]{\lineheight{1.25}\smash{\begin{tabular}[t]{l}$\rho$\end{tabular}}}}%
    \put(0.0380173,0.00103694){\color[rgb]{0,0,0}\makebox(0,0)[lt]{\lineheight{1.25}\smash{\begin{tabular}[t]{l}$\hat{x}$\end{tabular}}}}%
    \put(0.71838274,0.26585218){\color[rgb]{0,0,0}\makebox(0,0)[lt]{\lineheight{1.25}\smash{\begin{tabular}[t]{l}$\hat{y}$\end{tabular}}}}%
    \put(0.33525796,0.68772146){\color[rgb]{0,0,0}\makebox(0,0)[lt]{\lineheight{1.25}\smash{\begin{tabular}[t]{l}$\hat{z}$\end{tabular}}}}%
    \put(0,0){\includegraphics[width=\unitlength,page=2]{spherical_coords.pdf}}%
  \end{picture}%
\endgroup%

    \caption{Graphical depiction of the spherical parameterisation of an initial condition. The light green `cone' represents an example hyperplane of fixed $\phi$.}
    \label{f:mappings:spherical_coordinates}
\end{subfigure}\hfill%
\begin{subfigure}{0.45\textwidth}
    \centering
    \def\svgwidth{\linewidth}
%% Accompanies image file 'cartesian_coords.pdf' (pdf, eps, ps)
\begingroup%
  \makeatletter%
  \providecommand\color[2][]{%
    \errmessage{(Inkscape) Color is used for the text in Inkscape, but the package 'color.sty' is not loaded}%
    \renewcommand\color[2][]{}%
  }%
  \providecommand\transparent[1]{%
    \errmessage{(Inkscape) Transparency is used (non-zero) for the text in Inkscape, but the package 'transparent.sty' is not loaded}%
    \renewcommand\transparent[1]{}%
  }%
  \providecommand\rotatebox[2]{#2}%
  \newcommand*\fsize{\dimexpr\f@size pt\relax}%
  \newcommand*\lineheight[1]{\fontsize{\fsize}{#1\fsize}\selectfont}%
  \ifx\svgwidth\undefined%
    \setlength{\unitlength}{595.27559055bp}%
    \ifx\svgscale\undefined%
      \relax%
    \else%
      \setlength{\unitlength}{\unitlength * \real{\svgscale}}%
    \fi%
  \else%
    \setlength{\unitlength}{\svgwidth}%
  \fi%
  \global\let\svgwidth\undefined%
  \global\let\svgscale\undefined%
  \makeatother%
  \begin{picture}(1,0.7047619)%
    \lineheight{1}%
    \setlength\tabcolsep{0pt}%
    \put(0,0){\includegraphics[width=\unitlength,page=1]{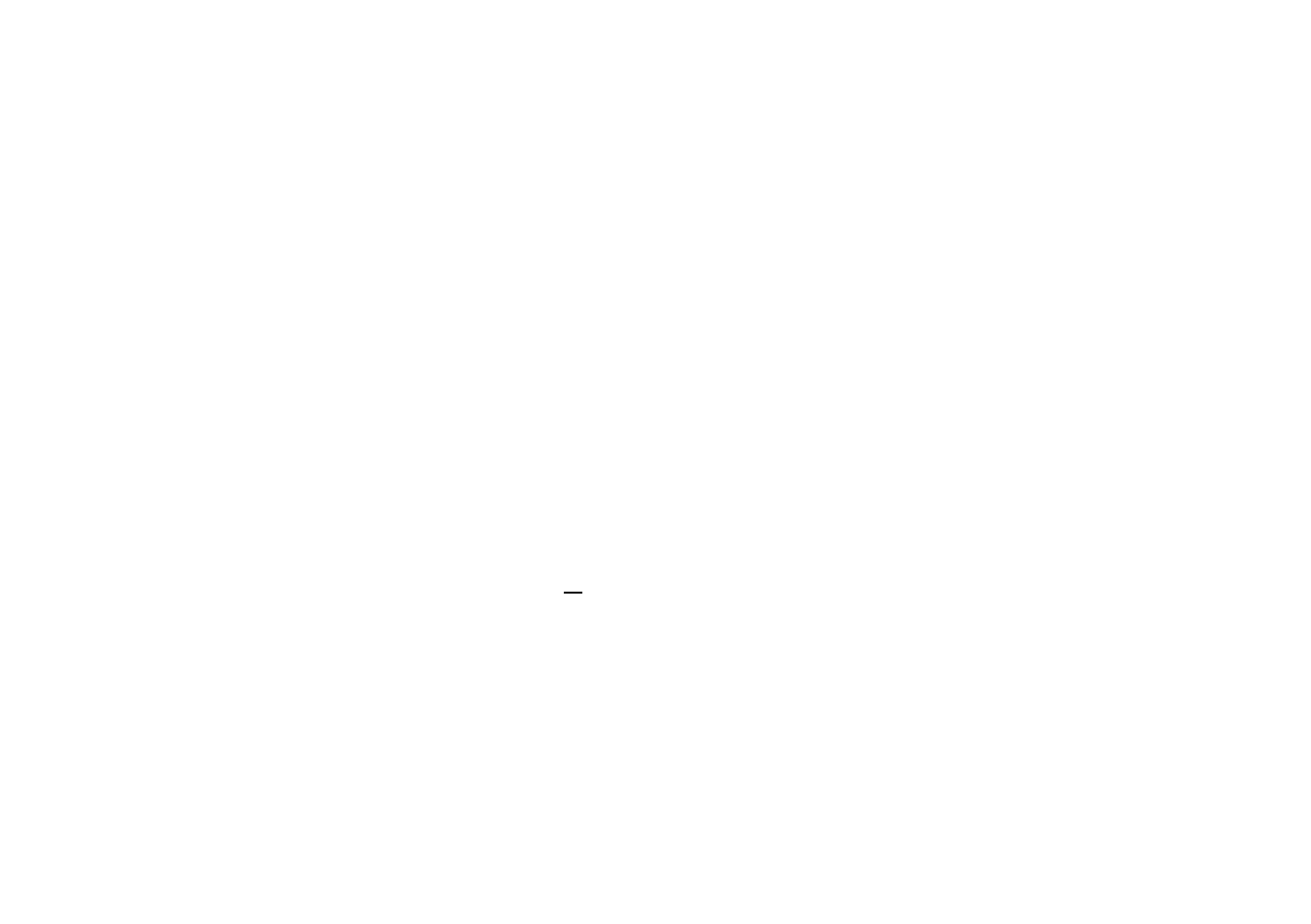}}%
    \put(0.1299782,0.00846617){\color[rgb]{0,0,0}\makebox(0,0)[lt]{\lineheight{1.25}\smash{\begin{tabular}[t]{l}$\hat{x}$\end{tabular}}}}%
    \put(0.81827679,0.24817548){\color[rgb]{0,0,0}\makebox(0,0)[lt]{\lineheight{1.25}\smash{\begin{tabular}[t]{l}$\hat{y}$\end{tabular}}}}%
    \put(0.43068475,0.63004876){\color[rgb]{0,0,0}\makebox(0,0)[lt]{\lineheight{1.25}\smash{\begin{tabular}[t]{l}$\hat{z}$\end{tabular}}}}%
    \put(0,0){\includegraphics[width=\unitlength,page=2]{cartesian_coords.pdf}}%
    \put(0.81750728,0.5436064){\color[rgb]{0,0,0}\makebox(0,0)[lt]{\lineheight{1.25}\smash{\begin{tabular}[t]{l}$z = c$\end{tabular}}}}%
    \put(0.81270704,0.41758702){\color[rgb]{0,0,0}\makebox(0,0)[lt]{\lineheight{1.25}\smash{\begin{tabular}[t]{l}$z = b$\end{tabular}}}}%
    \put(0.81270704,0.31120083){\color[rgb]{0,0,0}\makebox(0,0)[lt]{\lineheight{1.25}\smash{\begin{tabular}[t]{l}$z = a$\end{tabular}}}}%
  \end{picture}%
\endgroup%

    \caption{Graphical depiction of the Cartesian parameterisation of an initial condition. The light green planes represents examples of different hyperplanes of fixed $z$.}
    \label{f:mappings:cartesian_coordinates}
\end{subfigure}\hfill%
\caption{The parameterisations explored in this paper \revadd{to define the submanifolds}. For orbital elements, the embedding is defined by fixing the remaining orbital elements. For spherical and Cartesian coordinates, the same unique velocity is associated with each point in space as would be obtained from the orbital element transformation at that point.}
\end{figure}

The final transformation examined in this paper defines an initial condition in the non-rotating frame using Cartesian coordinates $\bm{x} = \left(x,\,y,\,z\right)$ with hyperplanes given by fixed values of $z$ (Figure \ref{f:mappings:cartesian_coordinates}). It is investigated here since it is conceptually simpler to visualise than the previous two approaches, but has not yet been investigated in the literature. 

The velocity to complete the full phase space is specified in the same manner as for spherical coordinates, following the procedure from Equation \ref{eq:velocitydefinition} onward. Overall, this transformation is the $\Psi-\Pi$ pair
\begin{gather}
\Psi_C: 
    \begin{Bmatrix}
        x\\y\\z
    \end{Bmatrix}_I \mapsto
    \begin{Bmatrix}
    x\\y\\z\\\dot{x}\\\dot{y}\\\dot{z}
    \end{Bmatrix}_I \mapsto
    \begin{Bmatrix}
    x\\y\\z\\x^\prime\\y^\prime\\z^\prime
    \end{Bmatrix}_R\\
\Pi_C: \begin{Bmatrix}\label{eq:picart}
    x\\y\\z\\x^\prime\\y^\prime\\z^\prime
    \end{Bmatrix}_R \mapsto
    \begin{Bmatrix}
    x\\y\\z\\\dot{x}\\\dot{y}\\\dot{z}
    \end{Bmatrix}_I \mapsto
    \begin{Bmatrix}
        x\\y\\z
    \end{Bmatrix}_I.
\end{gather}

\section{Results}

The results of the application of LCS to the Sun-Mars ER3BP are presented here. We first analyse in detail the LCS found for the Sun-Mars ER3BP using the orbital elements transformation and explain how we reconstruct the surface, then present results on the effect of orbit transformation and the choice of units that are beneficial for the user when numerically computing LCS.

\subsection{The LCS Structure}

\begin{figure}
    \centering
    \begin{subfigure}{0.48\textwidth}
        \def\svgwidth{\linewidth}
        {\tiny
        \includegraphics[width=\linewidth]{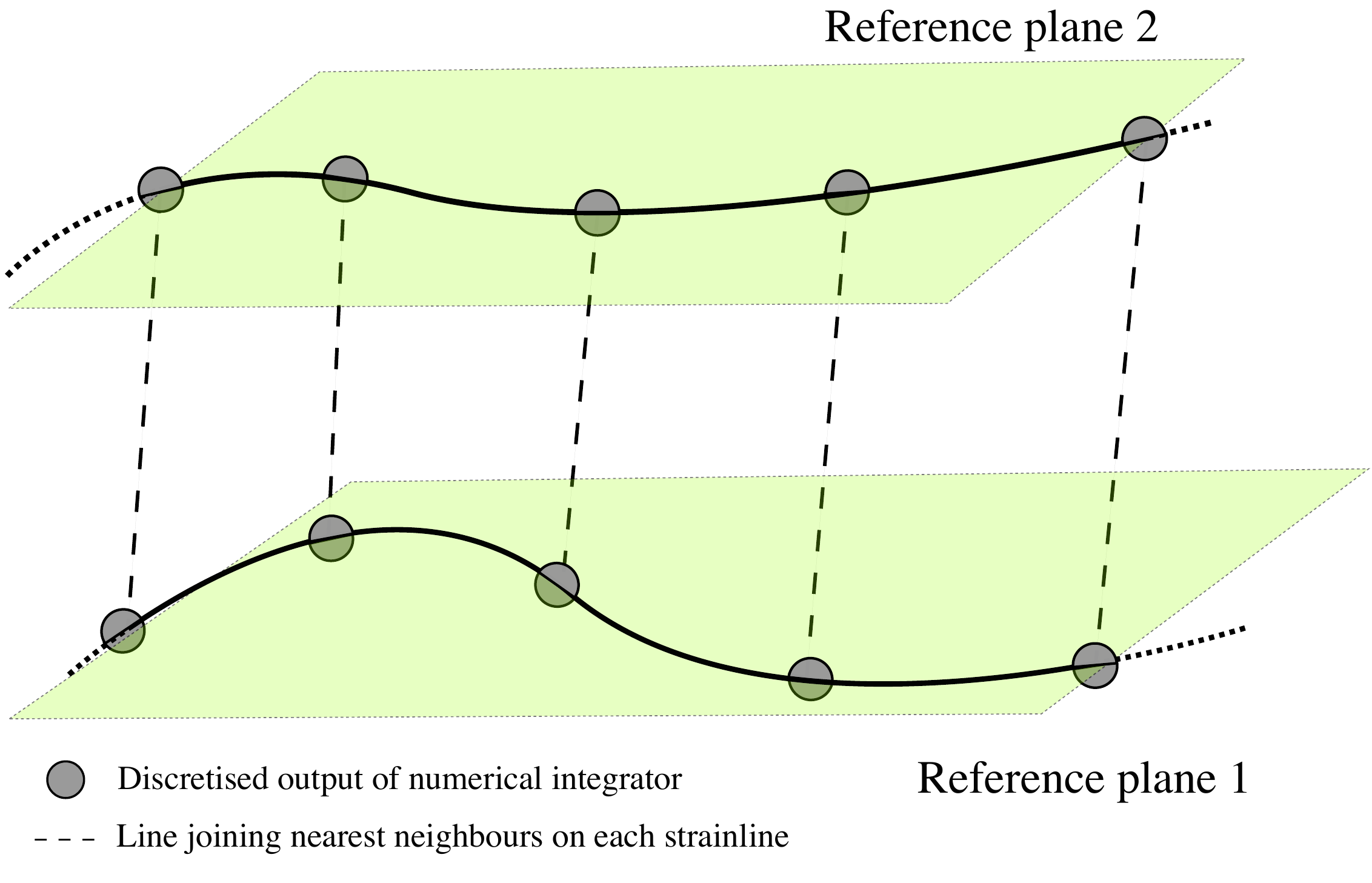}
        }
        \caption{The first step in reconstructing the full LCS structure is finding the nearest neighbours to points on adjacent reference planes by computing the Euclidean distance between them in parameter space.}
        \label{f:lcs_reconstruction_first_step}
    \end{subfigure}\hfill%
    \begin{subfigure}{0.48\textwidth}
        \includegraphics[width=\linewidth]{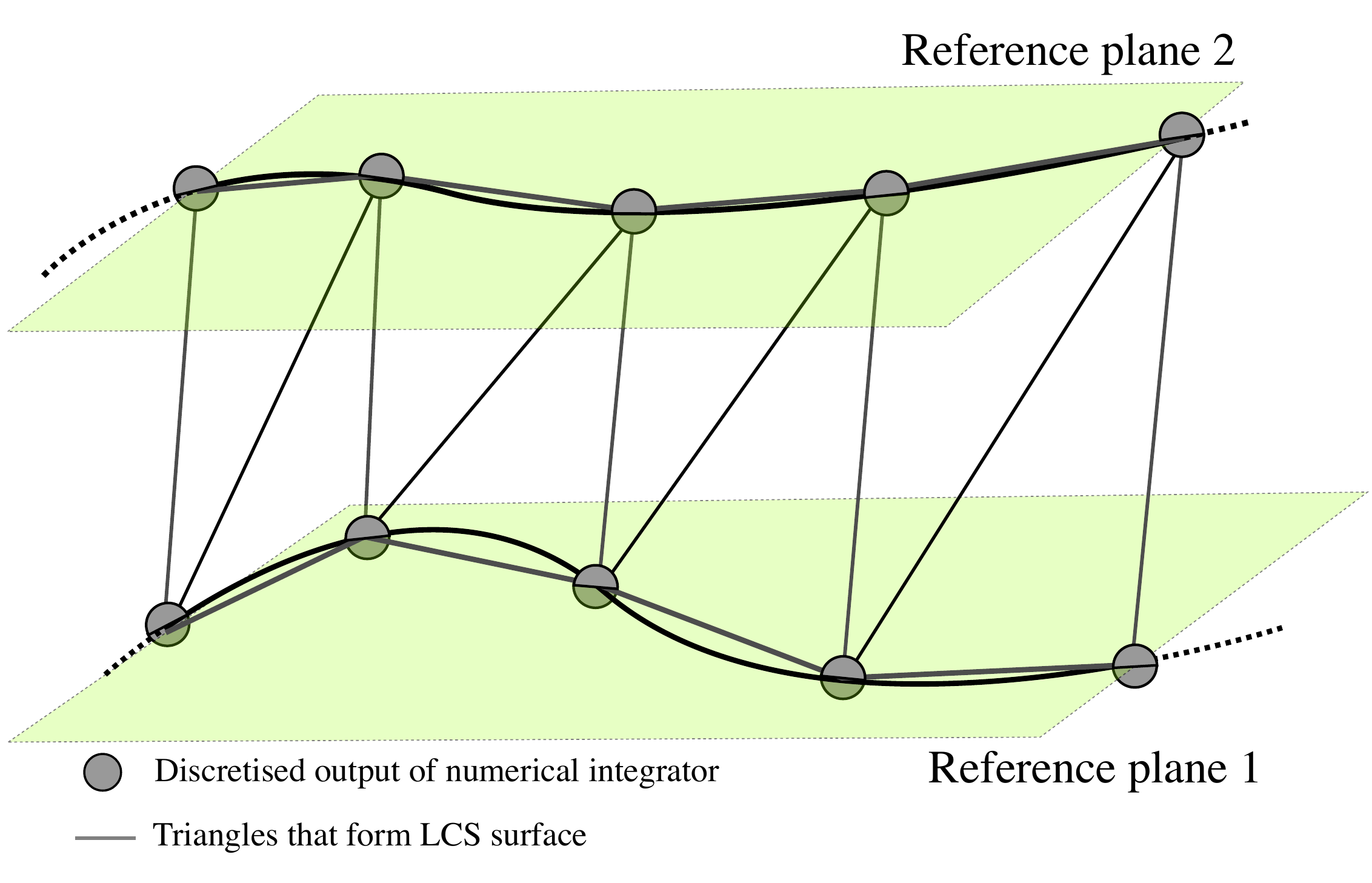}
        \caption{Triangles are formed between nearest neighbours in each reference plane, which can be written directly to a STL file and displayed and manipulated as a surface.}
        \label{f:lcs_reconstruction_second_step}
    \end{subfigure}
    \caption{Diagrams to show how the full LCS surface is reconstructed by interpolating between strainlines on neighbouring reference planes.}
\end{figure}

In this example, we analyse the Sun-Mars ER3BP from an initial integration time of $t_0 = \nu_0= 0$ to a final integration time of $T = \nu = 2\pi$ (1 full Martian year). $\Psi$ and $\Pi$ are the orbital elements transformation $\Psi_{OE}$ and $\Pi_{OE}$ and the helicity threshold $\alpha = 10^{-9}$. The set of reference planes $\mathcal{S}$ are defined such that $$\mathcal{S}_\text{OE} = \left\{r_p\in\left[r, r_s\right],\,\omega\in\left[0,2\pi\right],i\in\left[-85^\circ,\,-75^\circ,\dots,85^\circ\right]\right\}.$$ The variables $r = 1.641\times 10^{-5}$ and $r_s = 0.00513$ here are the radius and the Hill sphere of Mars in the scaled units of the non-rotating frame, respectively. 
A $100\times 2500$ grid in $r_p$ and $\omega$ is used on each plane in $\mathcal{S}$ to identify seed points. The motivation behind choosing this grid is elaborated later.

We produce the full structure of the LCS as a STL file, which is a common file format in 3D modelling applications. The STL file format contains a sequence of sets of three vertices that form triangles, and the collection of faces of the triangles are then displayed as a full structure by freely-available viewers, such as Blender\footnotemark[1] or Meshlab\footnotemark[2]. 

\footnotetext[1]{\url{https://blender.org}}
\footnotetext[2]{\url{https://meshlab.net}}

Since this file format is relatively straightforward to create, we create the STL file and thus the full structure using a triangulation algorithm. We first separate the full structure into a collection of strainlines on each reference plane in parameter space and, by leveraging that we expect there to be a nearby strainline in the neighbouring reference plane, we find the nearest neighbour to each point on a strainline in the neighbouring reference plane (Figure \ref{f:lcs_reconstruction_first_step}). This is performed by considering the Euclidean distance between points in adjacent reference planes in the parameter space of the submanifold.

Now the two points that should be interpolated between are known, a triangle is formed using the original point, its nearest neighbour in the next reference plane, and a point next to the original point in the same strainline (Figure \ref{f:lcs_reconstruction_second_step}). This procedure is then repeated for every point in every strainline in the reference plane, and then repeated across all reference planes. By doing so, we are able to construct a triangulation of the full structure of the LCS for use in visualisation software.

\begin{figure}
    \centering
    \begin{subfigure}{0.45\textwidth}
        \includegraphics[width=\linewidth]{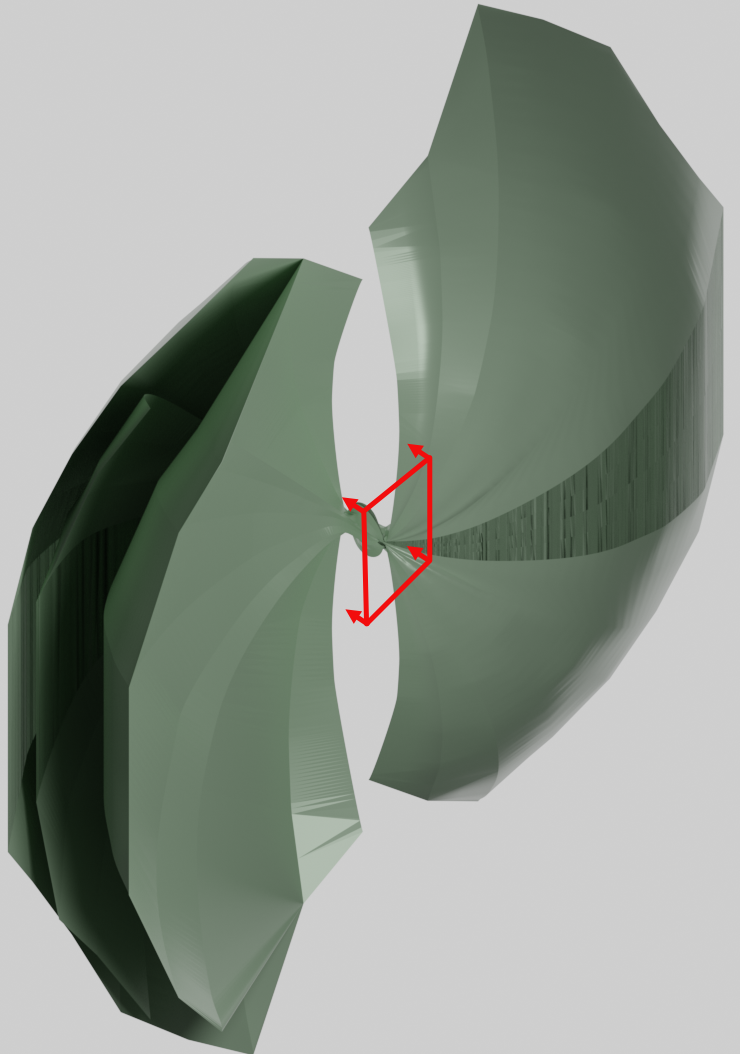}
        \caption{The full structure of the LCS obtained, which is comprised of an inner section and two large `arms' which emanate from the inner section. The region in red is the point-of-view of the section plot shown in more detail in Figure \ref{f:inner_lcs_section}.}
        \label{f:full_lcs_structure}
    \end{subfigure}\hfill
    \begin{subfigure}{0.45\textwidth}
        \includegraphics[width=\linewidth]{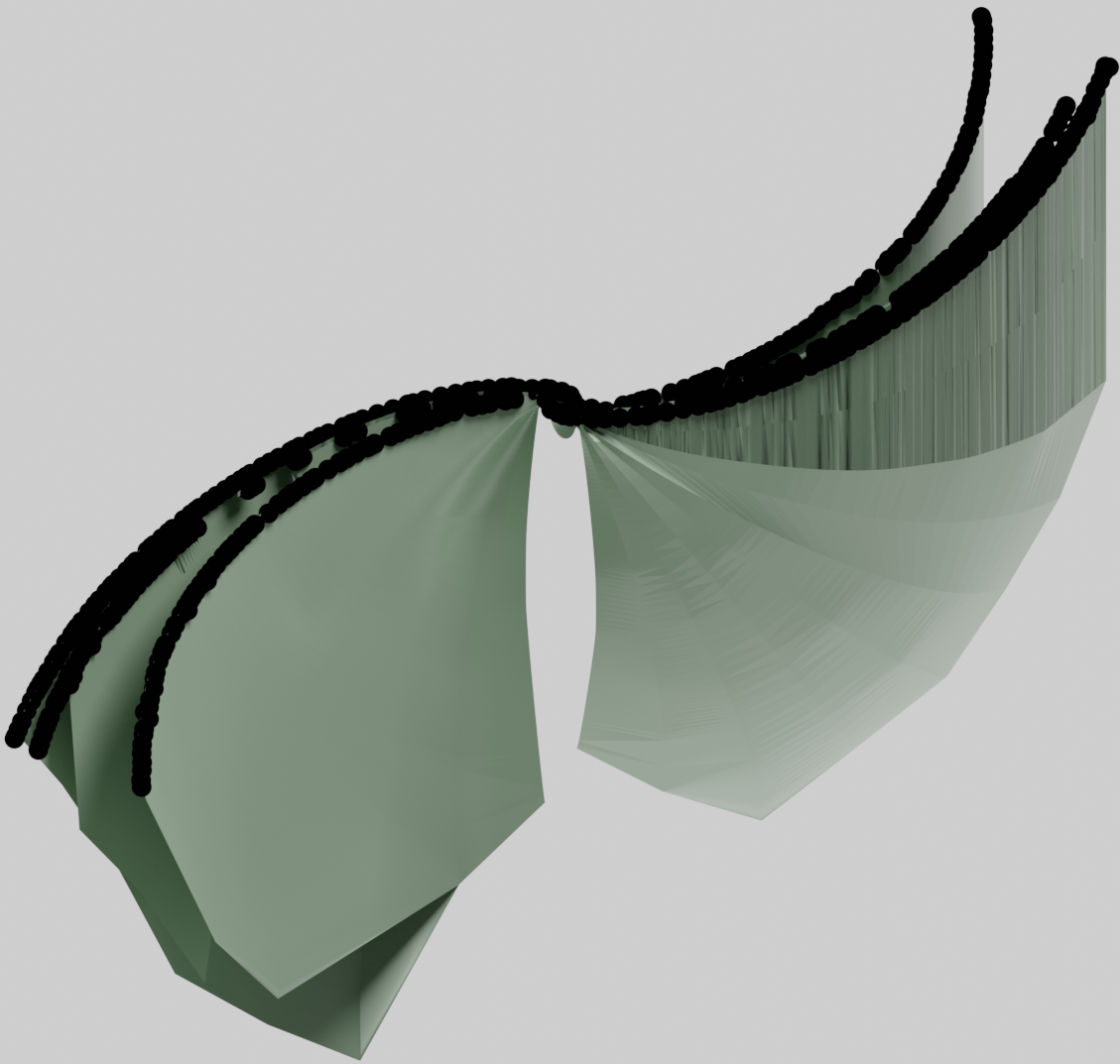}
        \caption{Section view of one of the hyperplanes of fixed inclination with points that form the strainlines shown in black to highlight more clearly how the surface is reconstructed.}
        \label{f:strainline_structure}
    \end{subfigure}\hfill%
    \begin{subfigure}{0.45\textwidth}
        \includegraphics[width=\linewidth]{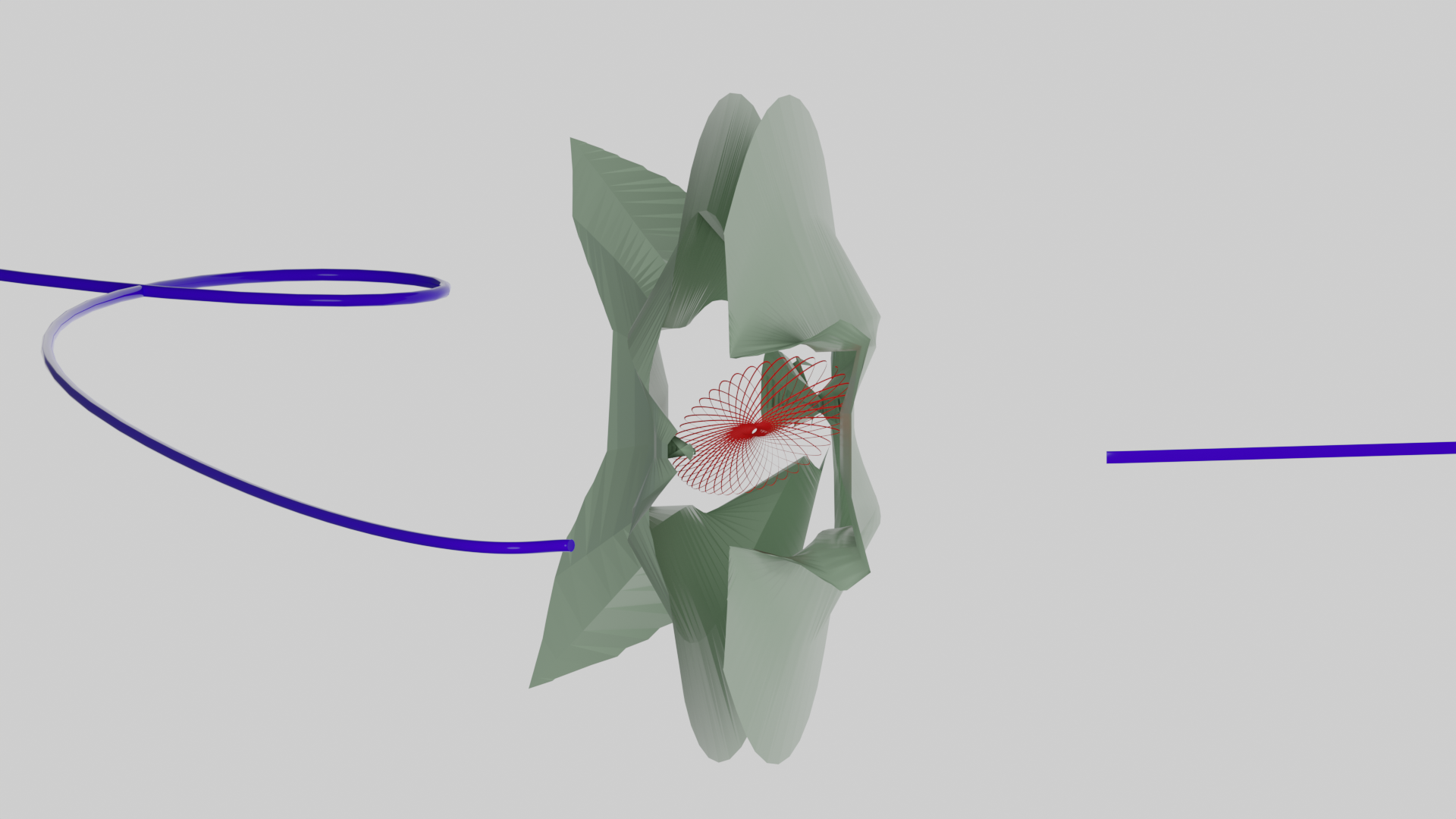}
        \caption{Inner section of the LCS (highlighted in red in Figure \ref{f:full_lcs_structure}) with the `arms' removed and three sample trajectories overlaid to show how it is acting as a separatrix. The trajectory taken from inside the LCS (red) remains orbiting Mars while the two trajectories that begin nearby but outside the inner section (blue) both immediately escape.}
        \label{f:inner_lcs_section}
    \end{subfigure}\hfill
    \begin{subfigure}{0.45\textwidth}
        \includegraphics[width=\linewidth]{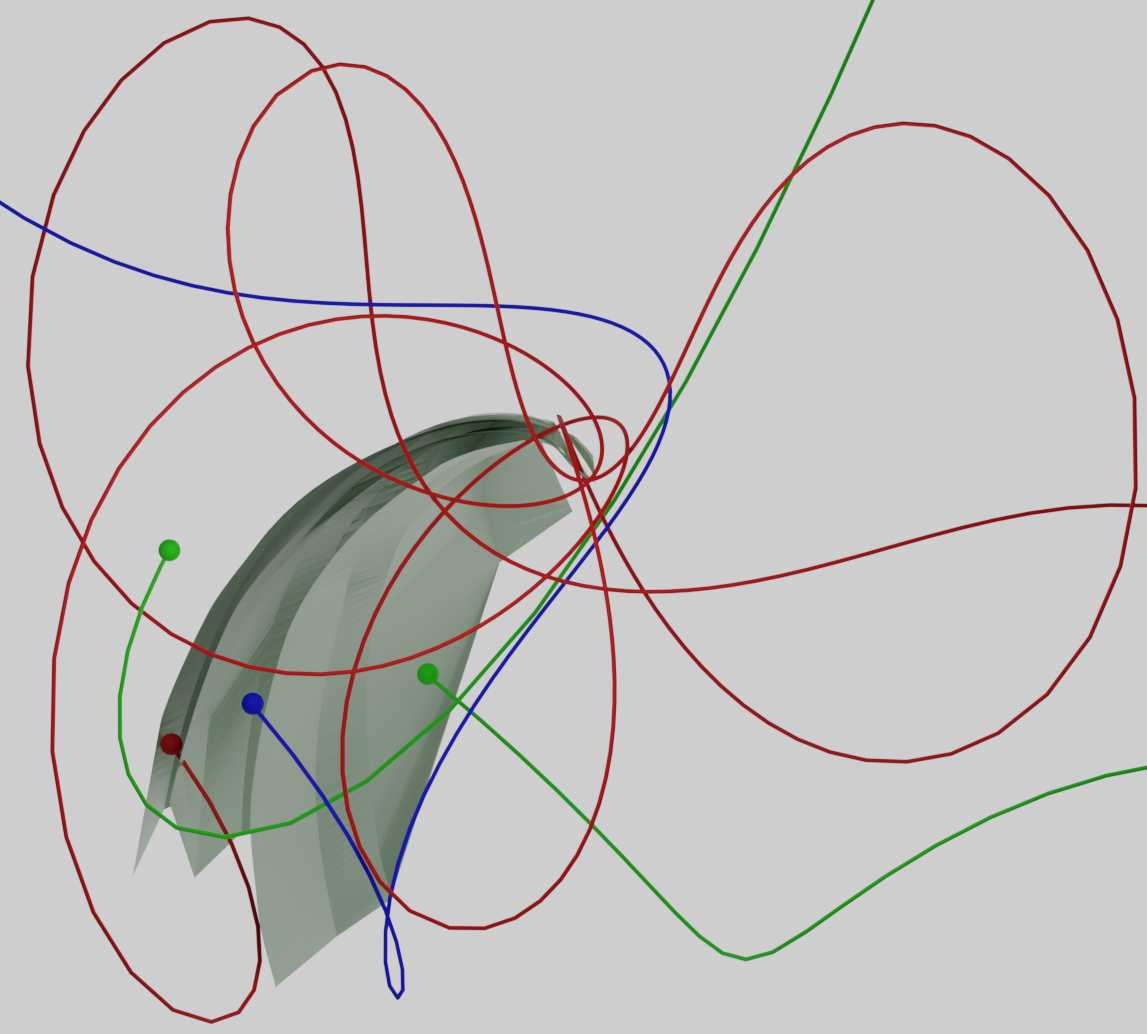}
        \caption{Trajectories in the synodic frame of points taken from different parts of the LCS arms. Each trajectory is qualitatively different and the LCS is again acting as a separatrix.}
        \label{f:lcs_arms}
    \end{subfigure}
    \caption{A series of representative renderings of the LCS structure corresponding to the Sun-Mars ER3BP and reference hyperplanes outlined in the main text. How the LCS is constructed from the strainlines is shown in more detail, as well as how the LCS is acting a separator between qualitatively different behaviour.}
    \label{f:lcs_structures}
\end{figure}

The full, interpolated repelling LCS structure from the above procedure is given in Figure \ref{f:full_lcs_structure}.  To show more clearly how the LCS is constructed from the strainlines on the reference hyperplanes, an example set of strainlines is shown on a section view of the full LCS in Figure \ref{f:strainline_structure}. The LCS is formed of two distinct sections: the `inner' portion that encloses Mars (Figure \ref{f:inner_lcs_section}) and the two extending arms that emanate from the inner structure (Figure \ref{f:lcs_arms});
these two regions are now investigated in more detail. 

In Figure \ref{f:inner_lcs_section} the inner structure is presented with the extending arms removed and example trajectories corresponding to some initial conditions inside and outside of the LCS are shown. In red we show a generic trajectory that begins inside the inner section, which continually orbits Mars over the integration period considered. Mars lies at the centre of the inner section. In blue are two trajectories that begin nearby but outside of the inner section, which both immediately escape from Mars' sphere of influence. The LCS here is acting as separator between trajectories which have qualitatively very different behaviour. This matches the expected structure from investigations into ballistic capture where desired behaviour is defined beforehand \citep{Luo2015}, but the insight here is obtained without any \textit{a priori} knowledge.

A more detailed presentation of one of the two `arms' is given in Figure \ref{f:lcs_arms}, again with example trajectories highlighted and the respective initial conditions indicated by spheres. In green are two trajectories which begin outside of the LCS structure and are qualitatively similar in immediately escaping the sphere of influence of Mars, as was the case in Figure \ref{f:inner_lcs_section}. More significantly, the remaining blue and red trajectories that are taken from different parts of the arms are also qualitatively different to both themselves and the two green trajectories outside of the LCS, highlighting how effectively the LCS is partitioning dynamical behaviour. In general, these `arms' separate orbits which complete a different number of revolutions from each other, including those which immediately escape from Mars. This again matches expectations from the literature \citep{Luo2015}, which identifies these as `stable sets'.

\subsection{Effect of the choice of orbit transformation and integration time}

In the previous Subsection, we showed the result of DA-LCS to the Sun-Mars ER3BP with a single choice of transformation $\Psi_{OE}$ and $\Pi_{OE}$ and a single prescribed time period of investigation. However, there are many free choices available to the mission designer when using LCS, such as the initial and final times to consider, and the choices of $\Pi$ and $\Phi$. In this Subsection we present the effect of the choice of orbit transformation and integration time on the helicity field found. The structure of the LCS can then be inferred by regions of consistently low helicity, which are coloured dark blue in the plots. An advantage of DA-LCS is that it makes the implementation of other transformations straightforward via automatic differentiation. 

\begin{sidewaysfigure}
    \begin{subfigure}{0.33\textwidth}
        \includegraphics[width=\linewidth]{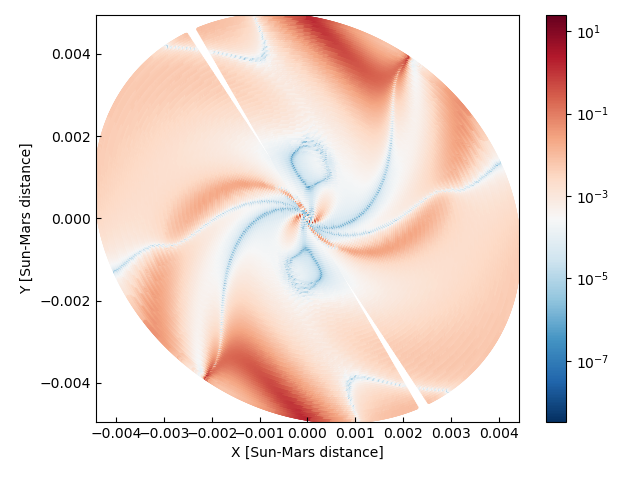}
        \caption{Orbital elements transformation.}
        \label{f:matrix:oes50}
    \end{subfigure}\hfill%
    \begin{subfigure}{0.33\textwidth}
        \includegraphics[width=\linewidth]{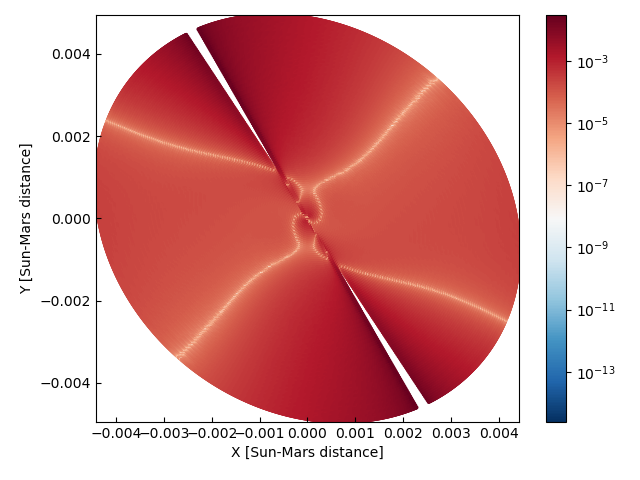}
        \caption{Spherical transformation.}
        \label{f:matrix:spherical50}
    \end{subfigure}%
    \begin{subfigure}{0.33\textwidth}
        \includegraphics[width=\linewidth]{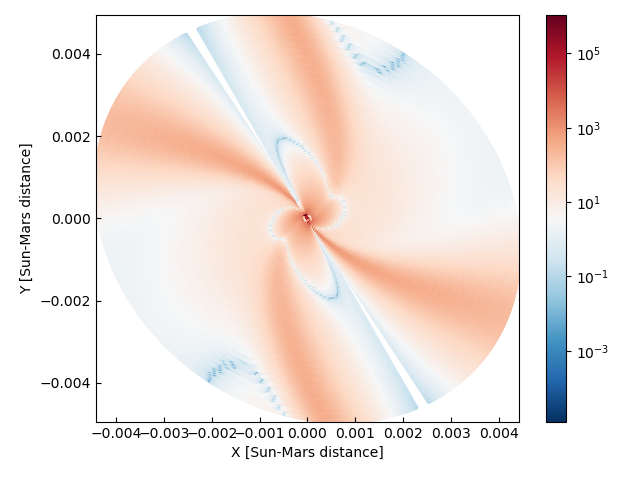}
        \caption{Cartesian transformation.}
        \label{f:matrix:cartesian50}
    \end{subfigure}
    \begin{subfigure}{0.33\textwidth}
        \includegraphics[width=\linewidth]{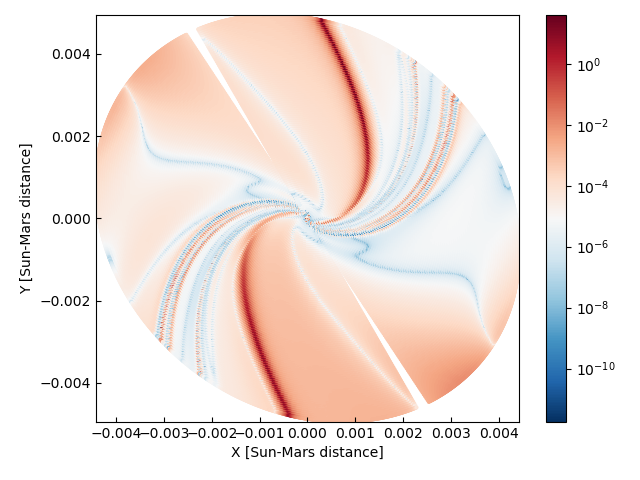}
        \caption{Orbital elements transformation.}
        \label{f:matrix:oes250}
    \end{subfigure}\hfill%
    \begin{subfigure}{0.33\textwidth}
        \includegraphics[width=\linewidth]{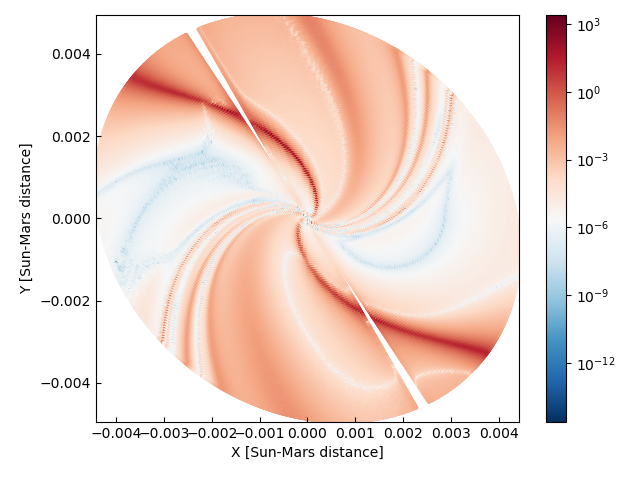}
        \caption{Spherical transformation.}
        \label{f:matrix:spherical250}
    \end{subfigure}%
    \begin{subfigure}{0.33\textwidth}
        \includegraphics[width=\linewidth]{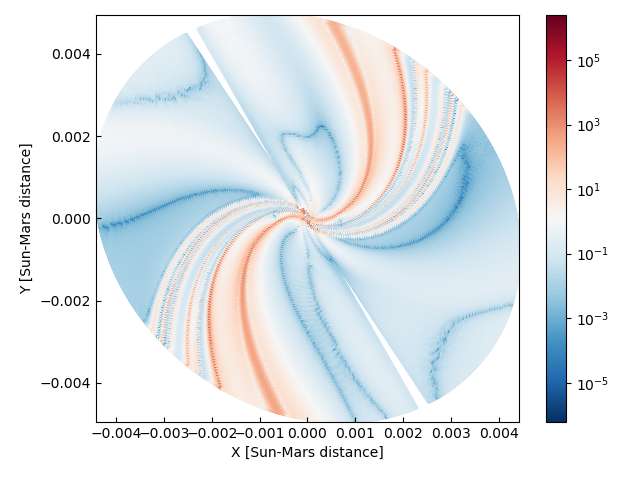}
        \caption{Cartesian transformation.}
        \label{f:matrix:cartesian250}
    \end{subfigure}%%========================================
    \caption{The effect of integration time on the helicity field for the ER3BP. The initial true anomaly is $t_0 = 0$ and the final true anomaly is either $T = \pi/4$ ($68$ days, top row), or $T=2\pi$ (one Martian year, bottom row). In general, sufficient time is required to remove the effect of the transformations on the final derivatives. Regions of low helicity (dark blue) imply the location of an LCS.}
    \label{f:matrix}
\end{sidewaysfigure}

We compare each $\Psi-\Pi$ pair introduced in Section \ref{sec:initialconditionstransfermap} for two integration times. The integration starts at $t_0 = \nu_0 = 0$ and ends at the true anomaly corresponding to $T = \nu = \pi / 4$ ($68$ days) and $T = 2\pi$ (1 Martian year) on the same set of initial conditions, to enable a fair comparison. The grid of initial conditions is chosen to be those which lie on the $i=35^\circ$ hyperplane in the orbital elements transformation, again sampled on a $100\times2500$ grid with the same parameters as used in the construction of the full LCS. For each point on the hyperplane in the orbital elements transformation, we also compute the equivalent Cartesian position. This Cartesian position forms the initial condition for the Cartesian transformation, and converting this Cartesian position to spherical coordinates gives the equivalent initial condition in the spherical transformation. In this way, we can ensure that each transformation samples the same points to ensure a valid comparison. \revadd{However,
by choosing to replicate the points on the $i=35^\circ$ plane exactly we will necessarily pass through the line of nodes, at which point the velocity attachment for spherical and Cartesian coordinates is ill-defined (Section \ref{sec:sphparam}). To avoid this we do not evaluate points within $2^\circ$ of the line of nodes. When using hyperplanes in $\phi$ or $z$, as discussed in Sections \ref{sec:sphparam} and \ref{sec:cartesianparam}, this problem does not arise.}

Figure \ref{f:matrix} presents the helicity fields for each of the transformations for both integration times.
We find that the integration time from $t_0$ to $T$ must be chosen to be sufficiently large to allow the time-dependent derivatives associated with the dynamical system in $\nf$ to dominate the time-independent derivatives associated with the transformations $\Psi$ and $\Pi$. For $T = \pi/4$, sufficient time has not passed for the derivatives associated with the dynamics to dominate, and the result is inconsistent between the different transformations. For $T = 2\pi$, however, the results are consistently representing the behaviour of the dynamical system. We find that $T = \pi$ is sufficient to ensure domination of the dynamics for all points in the domain for the examples shown here, though we note that many of the points farthest from Mars have orbital periods of thousands of days and much lower integration times are required for points closer to Mars.

Since the derivatives of the dynamics dominate the derivatives associated with the transformations after sufficient time, we would expect that each transformation yields similar helicity fields. The helicity fields signal qualitatively the same points as LCS, identifying the arms and the inner section as among the lowest helicity points for each transformation, despite differences in the actual value of helicity. 

\begin{figure}
    \centering
    \begin{subfigure}{0.48\textwidth}
        \includegraphics[width=\textwidth]{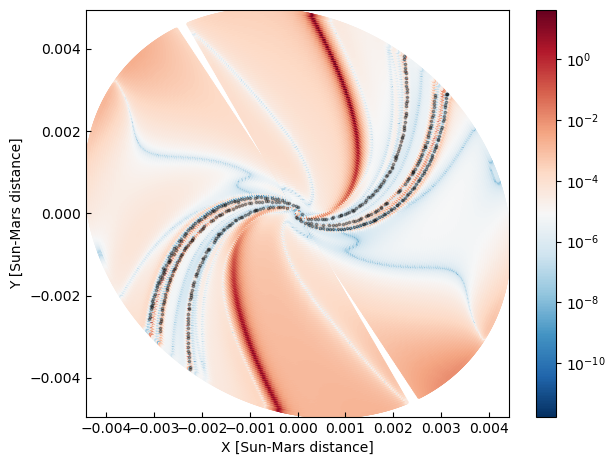}
        \caption{Orbital elements transformation.}
        \label{f:zoomed_helicity:oes}
    \end{subfigure}\hfill
    \begin{subfigure}{0.48\textwidth}
        \includegraphics[width=\textwidth]{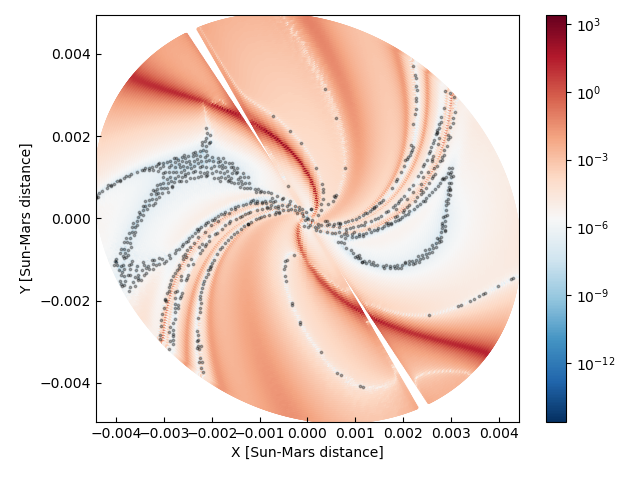}
        \caption{Spherical coordinates transformation.}
        \label{f:zoomed_helicity:spherical}
    \end{subfigure}\hfill
    \begin{subfigure}{0.48\textwidth}
        \includegraphics[width=\textwidth]{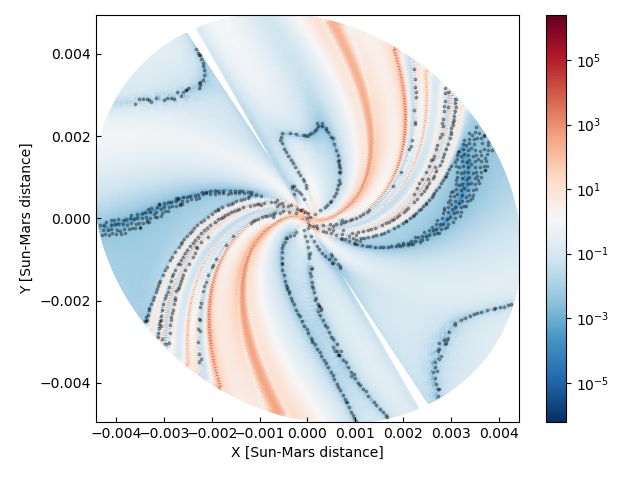}
        \caption{Cartesian transformation.}
        \label{f:zoomed_helicity:cartesian}
    \end{subfigure}
    \caption{\revadd{Helicity fields for each of the transformations introduced previously with the lowest points in the helicity field overlaid. The integration time is $2\pi$ (1 Martian year.)}}
    \label{f:zoomed_helicity}
\end{figure}

\revadd{Examining the quantitative agreement of the transformations from visual inspection of Figure \ref{f:matrix} is difficult since the width of the low-helicity regions is only approximately $40$ km. To highlight such narrow regions, in Figure \fixme{\ref{f:zoomed_helicity}} we overlay the lowest-helicity points in each field. Accurate sampling transverse to this very narrow region to ensure we find the minimum helicity points motivated the use of the asymmetric $100\times 2500$ grid choice in $r_p$ and $\omega$, with the denser grid in $\omega$ designed to densely sample initial conditions `across' the low-helicity regions. }

Figure \fixme{\ref{f:zoomed_helicity}} shows that the orbital elements and spherical transformations also quantitatively agree in both signalling the `arms' and inner section as the strictly lowest regions of helicity in each case, as is to be expected.
However, there is additional structure in the spherical transformation at this low helicity which does not agree with that predicted by the literature or the orbital elements case. With the Cartesian transformation this additional structure is signalled before \revadd{the structure of the arms is completed}. This structure is also discontinuous, despite the helicity being a continuous expression. By choosing a numerically challenging test case to benchmark DA-LCS on, we have exposed a numerical error that arises as a limit of floating-point arithmetic and manifests as `false positives' in the helicity field. We discuss this error and mitigating strategies in Section \ref{sec:inversion}.

We also find other structure at higher helicities in the orbital elements case, which arises as a result of the transformation also including some information on the final velocity in the three variables that parameterise the submanifold. Since obtaining $i$ and $\omega$ depends on the velocity, additional information on direction of escape and direction of final velocity is present compared to $\Pi_S$ and $\Pi_C$. In those, all final velocity information is discarded (Equations \ref{eq:pisph} and \ref{eq:picart}).
This additional structure extends that found in ballistic capture \citep{Luo2014ConstructingModel, Luo2015}, but these points do not feature in the full LCS: to be consistent with the original definition of the LCS, the helicity threshold is chosen to find only the most repelling structures. This yields the inner section and the `arms' before any additional structure.

\subsection{Implementation considerations}\label{sec:inversion}

A potential numerical pitfall exists when numerically constructing LCS in astrodynamics problems and is elaborated in this Subsection. We also provide guiding considerations on how to avoid these errors practically. We stress that since DA gives derivatives accurate to floating-point precision, the numerical difficulties given below are not as a result of DA-LCS, any particular software package or the mathematical definition of an LCS. Rather, they are a result of the finite precision of floating-point numbers and the large derivatives that arise in numerically challenging test cases.

Since computers approximate the field of real numbers using floating-point numbers, and these floating-point numbers have finite storage associated with them, there is a limit on the precision of numbers they can represent. Double-precision numbers have $64$ bits of storage, which provides a maximum of 16 significant digits that can be represented accurately \citep{Muller2010,IEEE754standard}. Any error arising from this finite-precision representation compared to the evaluation in real numbers is known as round-off error \citep{lantzQuantitativeEvaluationNumerical1971}.

Round-off error can occur in the evaluation of the helicity and construction of the LCS when $\nf$, $\cgst$ or $\nabla\bm{\zeta}_n$ become sufficiently ill-conditioned that the ratio of maximum to minimum absolute values of entries in those quantities approach $10^{16}$ in any relevant orders of the expansion. At this point, when the user requests a manipulation of these quantities, information on the smaller numbers is lost as a result of the finite precision. This error presents as false positives (false low values) in the helicity field.

\begin{sidewaysfigure}
    \centering
    \begin{subfigure}{0.31\textwidth}
        \centering
        \includegraphics[width=\linewidth]{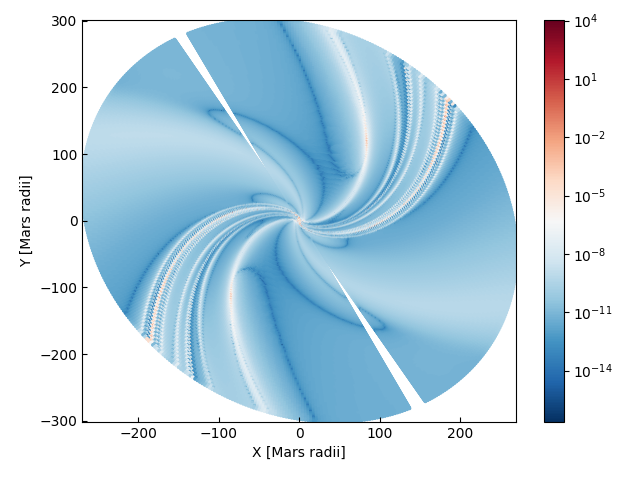}
        \caption{Orbital elements transformation, Mars radius.}
        \label{f:inversion:orbitalelementsR}
    \end{subfigure}\hfill%
    \begin{subfigure}{0.31\textwidth}
        \centering
        \includegraphics[width=\linewidth]{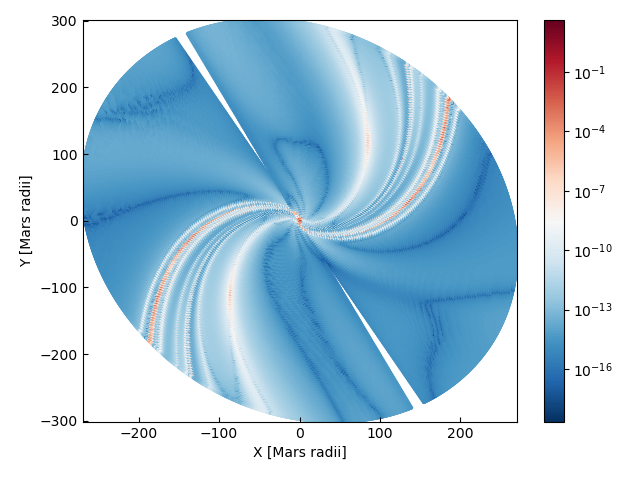}
        \caption{Spherical coordinates transformation, Mars radius.}
        \label{f:inversion:sphericalR}
    \end{subfigure}\hfill%
    \begin{subfigure}{0.31\textwidth}
        \centering
        \includegraphics[width=\linewidth]{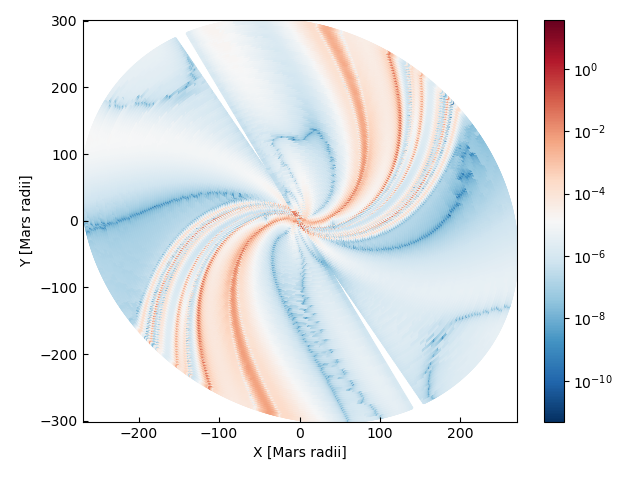}
        \caption{Cartesian coordinates transformation, Mars radius.}
        \label{f:inversion:cartesianR}
    \end{subfigure}
    \begin{subfigure}{0.31\textwidth}
        \centering
        \includegraphics[width=\linewidth]{no_line/no_line_oe_RP.png}
        \caption{Orbital elements transformation, Sun-Mars distance.}
        \label{f:inversion:orbitalelements10RP}
    \end{subfigure}\hfill%
    \begin{subfigure}{0.31\textwidth}
        \centering
        \includegraphics[width=\linewidth]{no_line/no_line_old_vel_helicity_field_spherical_RP.png}
        \caption{Spherical coordinates transformation, Sun-Mars distance.}
        \label{f:inversion:spherical10RP}
    \end{subfigure}\hfill%
    \begin{subfigure}{0.31\textwidth}
        \centering
        \includegraphics[width=\linewidth]{no_line/no_line_cartesian_RP.png}
        \caption{Cartesian coordinates transformation, Sun-Mars distance.}
        \label{f:inversion:cartesian10RP}
    \end{subfigure}
    \caption{\revadd{The helicity fields presented for each of the transformations using two unit lengths: Mars radius and Sun-Mars distance. We can aid the conditioning of relevant terms by scaling the unit length such that $\cgst$ is well-conditioned in parameterisations with some bounded terms.}}
    \label{f:inversion}
\end{sidewaysfigure}

We find that there are two mitigating strategies to prevent the ill-conditioning of these terms and prevent false positives in the helicity field in even the most challenging test cases. The first is to ensure the proper conditioning of relevant quantities by adjusting the magnitude of the terms of the orbit parameterisation. \revadd{In Figure \ref{f:inversion} we present the helicity fields computed for each of the transformations with two different unit lengths: Mars radius and Sun-Mars distance at Mars periapsis} to deliberately induce this round-off error. When using Mars radius as unit length for the orbital elements transformation (Figure \ref{f:inversion:orbitalelementsR}) and the spherical coordinates transformation (Figure \ref{f:inversion:sphericalR}) which are both formed of a radius term and two angles, the helicity field is `inverted', with low-helicity values in trajectories which all have qualitatively similar behaviour in immediately escaping, and higher values in the arms and inner section. A trajectory that immediately escapes from Mars will have a large final radius and associated derivatives, but with the angle terms still bounded in $\left[0,\,2\pi\right)$. The terms in $\cgst$ associated with the radius will thus dominate those associated with derivatives of the angles. When the helicity at this point is then evaluated, it becomes small not because of system dynamics, but because of a loss of precision of the arithmetic. Small numbers that would have otherwise contributed to the helicity have been lost to round-off error. \revadd{Condition numbers $\mathcal{C}$ for $\nabla\bm{
\zeta}_n$ of up to $10^{40}$ have been observed when using Mars radius as the unit length with the orbital elements transformation.}

However, this only occurs in regions of large final radius, i.e. points which immediately escape. Where the final radius is lower, such as in the arms where orbits complete several revolutions before escaping, the relevant terms are sufficiently well-conditioned and thus helicity values are sufficiently free of round-off errors that the `correct' order of magnitude of helicity is obtained at this point. Thus, lower helicities are present in regions of trajectories which all have qualitatively similar behaviour (immediately escaping) rather than in regions that separate the behaviour of trajectories, giving rise to the apparent inversion.

\revadd{We find we can aid the conditioning of relevant terms in transformations where some of the variables that parameterise the submanifold remain bounded, as is the case for angles in the orbital elements and spherical coordinates transformations. For these, the distance units should be chosen such that the entries of $\cgst$ are all of similar magnitude. This was achieved in this paper by using the Sun-Mars distance as unit length. Changing only the length unit in the orbital elements transformation from Mars radius in Figure \ref{f:inversion:orbitalelementsR} to the Sun-Mars distance in Figure \ref{f:inversion:orbitalelements10RP} reduces the condition number to approximately $\mathcal{C} = 10^{14}$ such that the helicity can be accurately represented and yields the expected result. We stress that these numerical effects are not specific to DA-LCS, any particular software package, or the mathematical definition of LCS. Rather, it is inherent to floating-point arithmetic and has been highlighted by the choice of a numerically particularly challenging test case.}

\revadd{Scaling units to ensure the Jacobian is well-conditioned is more difficult in systems where all variables that parameterise the submanifold are unbounded, such as the Cartesian transformation. Here, where trajectories typically escape in the direction of $x$ and $y$ but stay close to the plane in $z$, terms in $z$ can be dominated. Scaling the length unit uniformly will affect all terms equally and not change the condition number, which is above $10^{16}$ after even $68$ days. This causes the poor numerical resolution in the Cartesian transformation observed earlier: scaling all terms of the parameterisation does not affect round-off error in regions of escape, causing the additional structure in trajectories that immediately escape to persist.}

\revadd{The second mitigating strategy is to use a transformation that is naturally more robust to ill-conditioning. Even when using Sun-Mars distance as the unit length in spherical coordinates (Figure \ref{f:inversion:spherical10RP}) there remains sufficient ill-conditioning in the trajectories which escape ($\mathcal{C} \sim 10^{17}$) that false positives are present at the same helicity as the arms and inner section. However, for the orbital elements transformation the helicity can be evaluated accurately at all points in the field. We remark that we have seen satisfactory performance from spherical coordinates in less challenging test cases (see \citet{Tyler2022}) with values of $\Omega$ or $e$ that yield trajectories which escape less quickly. The numerical stability of orbital elements has been noted by other authors in similar situations (for example, \citet{Martin2022PeriodicNetworks}).}

\revadd{We therefore recommend against the use of the Cartesian transformation, except in isotropic systems or where one scales each term in the parameterisation non-uniformly using different unit lengths. In challenging test cases, we recommend the use of orbital elements to define initial conditions, which have produced usable insight even for the challenging problem parameters selected here. They also provide a more natural expression of defining initial conditions in the context of mission design. Nonetheless, other transformations may still yield better results depending on the system being studied.}

\revadd{We again stress that since DA provides derivatives accurate to machine precision, the above numerical difficulties are a result of double precision floating-point arithmetic. The use of quadruple precision (allowing up to 36 significant digits) would prevent round-off errors until approaching condition numbers of $10^{36}$. However, quadruple precision is not currently implemented natively on most CPUs, and software solutions are generally computationally very intensive and do not necessarily guarantee reproducibility between platforms \citep{IEEE754standard, Muller2010}.}

\section{Conclusion}\label{sec:conclusion}

We have presented the the results of the application of DA-LCS, a new numerical method suggested by the authors for the computation of Lagrangian Coherent Structures in three-dimensions. A full three-dimensional Lagrangian Coherent Structure in the Sun-Mars ER3BP has been computed accurately with DA-LCS, even in a numerically challenging test case designed to benchmark the method against. Detailed analysis of the LCS highlighted how it in this case acts as a separator between regions of qualitatively different behaviour, generalising the concept of stable and unstable manifolds. We discovered that sufficient integration time is required to ensure the LCS found reflects the system dynamics, and that care must be taken in defining initial conditions to avoid regions of particularly poor numerical numerical performance due to limitations in floating-point arithmetic. 
While the LCS was determined here for ER3BP, there is no obstacle to computing it in other systems provided the arithmetic can be readily replaced by DA operations; high-fidelity ephemeris models have been tested with DA-LCS and shown to yield similar insight as for the ER3BP case studied here.

We have several suggestions for future work. Firstly, we sampled the helicity on a $100\times 2500$ grid to find seed points for the strainlines. However, the major advantage of a LCS is that strainlines can be propagated forward once a single seed point is known, avoiding the computation of trajectories that do not otherwise correspond to the structure of the LCS. It would be a worthwhile pursuit to examine whether more sophisticated grid search strategies would reduce computation time, potentially exploiting a DA expansion of the helicity in combination with a gradient descent method to identify minima through local optimisation rather than grid search. Moreover, the extension of LCS to $n-$dimensions would avoid the need to define the dimension reductions introduced in this paper, and is numerically more simple to undertake than standard approaches if one follows the same methods used in DA-LCS. However, the interpretation and visualisation of $n-$dimensional structures remains a challenge, and so significant effort would be required to store, manipulate and visualise these structures.

To support this work and advance the use of LCS in astrodynamics, the source code for DA-LCS and the program used to produce this study will be made freely available under a GNU GPLv3 license.

% \backmatter

\bmhead{Acknowledgments}

The authors would like to acknowledge financial support from the EPSRC Centre for Doctoral Training in Next Generation Computational Modelling grant EP/L015382/1, and the use of the IRIDIS High Performance Computing Facility and associated support services at the University of Southampton.

\bibliographystyle{plainnat}
\bibliography{references_manual}


\end{document}